\documentclass[12pt]{JHEP3}

\usepackage{epsfig,latexsym,amssymb,cite}
\usepackage{amsmath}
\usepackage{graphicx}
\usepackage{bm}
\usepackage{slashed}

\newcommand{\be}{\begin{equation}}
\newcommand{\ee}{\end{equation}}
\newcommand{\bea}{\begin{eqnarray}}
\newcommand{\eea}{\end{eqnarray}}
\newcommand{\bem}{\begin{multline}}
\newcommand{\eem}{\end{multline}}
\newcommand{\beg}{\begin{gather}}
\newcommand{\eeg}{\end{gather}}

\newcommand{\nn}{\nonumber}

\newcommand{\as}{\alpha_s}

\def\eq#1{{Eq.~(\ref{#1})}}
\def\fig#1{{Fig.~\ref{#1}}}
\newcommand{\ben}{\begin{eqnarray*}}
\newcommand{\een}{\end{eqnarray*}}

\newcommand{\amu}{\alpha_\mu}

\newcommand{\Tr}{\text{Tr}}

\newcommand{\msbar}{\mu_{\overline {\text{MS}}}}
\def\peq#1{{(\ref{#1})}}
\newcommand{\sh}[1]{#1\hskip-7pt \diagup}

\newcommand{\wv}[1]{\boldsymbol{#1}}
\newcommand{\Sec}[1]{Sec.\ \ref{#1}}

\graphicspath{{Figs/}}

\title{Running Coupling Corrections to High Energy Inclusive Gluon Production}

\author{W. A. Horowitz,$^{1, 2}$ \ Yuri V. Kovchegov$^{\ 1}$ \\
\vspace{0.1in}

$^1$Department of Physics, The Ohio State University, Columbus,
OH 43210, USA \\

$^2$Department of Physics, The University of Cape Town, Rondebosch 7701, South Africa \\~~\\ 

E-mail addresses: \email{horowitz@mps.ohio-state.edu}, \email{yuri@mps.ohio-state.edu}

\vspace{0.1in}
}

\abstract{We calculate running coupling corrections for the
  lowest-order gluon production cross section in high energy hadronic
  and nuclear scattering using the BLM scale-setting prescription. In
  the final answer for the cross section the three powers of fixed
  coupling are replaced by seven factors of running coupling, five in
  the numerator and two in the denominator, forming a `septumvirate'
  of running couplings, analogous to the `triumvirate' of running
  couplings found earlier for the small-$x$ BFKL/BK/JIMWLK evolution
  equations. It is interesting to note that the two running couplings
  in the denominator of the `septumvirate' run with complex-valued
  momentum scales, which are complex conjugates of each other, such
  that the production cross section is indeed real. We use our
  lowest-order result to conjecture how running coupling corrections
  may enter the full fixed-coupling $k_T$-factorization formula for
  gluon production which includes non-linear small-$x$ evolution.}


\preprint{}

\begin{document}


\section{Introduction}

With the advent of the LHC era the physics of parton saturation/Color
Glass Condensate (CGC) \cite{Gribov:1984tu, Blaizot:1987nc,
  Mueller:1986wy, Mueller:1994rr, Mueller:1994jq, Mueller:1995gb,
  McLerran:1993ka, McLerran:1993ni, McLerran:1994vd, Kovchegov:1996ty,
  Kovchegov:1997pc, Jalilian-Marian:1997xn, Jalilian-Marian:1997jx,
  Jalilian-Marian:1997gr, Jalilian-Marian:1997dw,
  Jalilian-Marian:1998cb, Kovner:2000pt, Weigert:2000gi, Iancu:2000hn,
  Ferreiro:2001qy, Kovchegov:1999yj, Kovchegov:1999ua,
  Balitsky:1996ub, Balitsky:1997mk, Balitsky:1998ya, Iancu:2003xm,
  Weigert:2005us, Jalilian-Marian:2005jf} needs to produce more
quantitative predictions with higher accuracy. To achieve this goal
running coupling corrections to the
Jalilian-Marian--Iancu--McLerran--Weigert--Leonidov--Kovner (JIMWLK)
\cite{Jalilian-Marian:1997jx, Jalilian-Marian:1997gr,
  Jalilian-Marian:1997dw, Jalilian-Marian:1998cb, Kovner:2000pt,
  Weigert:2000gi, Iancu:2000hn, Ferreiro:2001qy} and
Balitsky--Kovchegov (BK) \cite{Balitsky:1996ub,
  Balitsky:1997mk,Balitsky:1998ya,Kovchegov:1999yj, Kovchegov:1999ua}
evolution equations have been calculated recently in
\cite{Kovchegov:2006vj,Balitsky:2006wa,Kovchegov:2006wf,Albacete:2007yr}
(see also \cite{Braun:1994mw,Levin:1994di}).  This led to a
significant improvement in the comparison of CGC predictions with data
from heavy ion and deep inelastic scattering (DIS) experiments
\cite{Albacete:2007sm,Albacete:2009fh,Kutak:2004ym}. However, to make
such comparisons theoretically consistent for observables like hadron
production in proton-proton ($pp$), proton-nucleus ($pA$) and
nucleus-nucleus ($AA$) collisions one needs to also include running
coupling corrections into the formulas for particle production.

While an exact analytic formula for gluon production in $AA$
collisions is still not known, we do know that in $pp$ and $pA$
collisions gluon production at the level of classical gluon fields and
leading-$\ln 1/x$ nonlinear quantum evolution is given by the
$k_T$-factorization formula
\cite{Kovchegov:2001sc,Braun:2000bh,Kharzeev:2003wz,Kovchegov:1998bi,Kovner:2006wr,Gribov:1984tu}:
\begin{align}\label{ktfact}
  \frac{d \sigma}{d^2 k_T \, dy} \, = \, \frac{2 \, \as}{C_F} \,
  \frac{1}{{\bm k}^2} \, \int d^2 q \, \phi_p ({\bm q}, y) \, \phi_{A}
  ({{\bm k} - \bm q}, Y-y).
\end{align}
Here $Y$ is the total rapidity interval of the collision, $C_F =
(N_c^2 -1)/2 N_c$, boldface variables denote two-component transverse
plane vectors ${\bm k} = (k^1 , k^2)$, and $\phi_p$, $\phi_A$ are the
unintegrated gluon distributions in the proton and the nucleus,
respectively, which are defined by
\cite{Kovchegov:2001sc,Kharzeev:2003wz,Jalilian-Marian:2005jf}
\begin{align}\label{ktglueA}
  \phi_A ({\bm k}, y) \, = \, \frac{C_F}{\as \, (2 \pi)^3} \, \int d^2 b \, 
d^2 r \, e^{- i {\bm k} \cdot {\bm r}} \ \nabla^2_r \, N_G ({\bm r},
{\bm b}, y)
\end{align}
and 
\begin{align}\label{ktgluep}
\phi_p ({\bm k}, y) \, = \, \frac{C_F}{\as \, (2 \pi)^3} \, \int d^2 b \, 
d^2 r \, e^{- i {\bm k} \cdot {\bm r}} \ \nabla^2_r \, n_G ({\bm r},
{\bm b}, y).
\end{align}
In \eq{ktglueA} the quantity $N_G ({\bm r}, {\bm b}, y)$ denotes the
forward scattering amplitude for a gluon dipole of transverse size
$\bm r$ with its center located at the impact parameter $\bm b$
scattering on a target nucleus with total rapidity interval $y$.
$N_G ({\bm r}, {\bm b}, y)$ can, in general, be found from the JIMWLK
evolution equation \cite{Jalilian-Marian:1997jx,
  Jalilian-Marian:1997gr, Jalilian-Marian:1997dw,
  Jalilian-Marian:1998cb, Kovner:2000pt, Weigert:2000gi, Iancu:2000hn,
  Ferreiro:2001qy}. In the large-$N_c$ limit it is related to the
quark dipole forward scattering amplitude on the same nucleus $N ({\bm
  r}, {\bm b}, y)$ by
\begin{align}\label{2NN}
  N_G ({\bm r}, {\bm b}, y) \, = \, 2 \, N ({\bm r}, {\bm b}, y) - N
  ({\bm r}, {\bm b}, y)^2,
\end{align}
where $N ({\bm r}, {\bm b}, y)$ can be found from the BK evolution
equation \cite{Balitsky:1996ub,
  Balitsky:1997mk,Balitsky:1998ya,Kovchegov:1999yj, Kovchegov:1999ua}
(see also \cite{Kovchegov:2008mk} for approximate ways of finding
$N_G$ beyond the large-$N_c$ limit without solving the JIMWLK equation).
The quantity $n_G ({\bm r}, {\bm b}, y)$ from \eq{ktgluep} is also a
gluon dipole amplitude, but taken in a dilute regime, where it is
found by solving the linear Balitsky-Fadin-Kuraev-Lipatov (BFKL)
evolution equation \cite{Bal-Lip,Kuraev:1977fs}.

In our notation the projectile or target is referred to as a `proton'
if the transverse momenta of interest in the problem are much larger
than the projectile's (target's) saturation scale, such that only
linear evolution is needed to describe such a projectile/target.
Conversely, if the saturation scale of the projectile/target is
comparable to the momentum scales of interest then we refer to this
projectile/target as a `nucleus'. With this notation, \eq{ktfact} is
not an assumption, but an exact answer which resulted in a non-obvious
way from a diagram resummation done in
\cite{Kovchegov:2001sc,Kovchegov:1998bi}.

As mentioned above, the running coupling corrections have been
calculated for the BFKL, BK, and JIMWLK evolution equations in
\cite{Kovchegov:2006vj,Balitsky:2006wa,Kovchegov:2006wf,Albacete:2007yr}
and for the initial conditions to these evolution equations in the
Appendix of \cite{Kovchegov:2007vf}, yielding a prescription of how to
find $N_G$ and $n_G$ in Eqs.\ (\ref{ktglueA}) and (\ref{ktgluep})
including running coupling corrections. However, there are three more
factors of strong coupling $\as$ in Eqs.\ (\ref{ktfact}),
(\ref{ktglueA}), and (\ref{ktgluep}) for which one needs to set the
scale. The first steps in this direction were taken in
\cite{Kovchegov:2007vf}, where it was shown that, in accordance with
conventional wisdom, in order to include running coupling corrections
in \eq{ktfact} one has to slightly re-define the observable: on top of
gluon production, one has to allow for contributions where the
would-be produced gluon splits into a collinear gluon-gluon (or
quark--anti-quark) pair. To ensure that the particles in the pair are
really collinear, one can put a bound on the virtuality of the gluon
splitting into a pair, by requiring that $k^2 <
\Lambda_\text{coll}^2$, where $k^\mu$ is the four-momentum of the
would-be produced gluon and $\Lambda_\text{coll}$ is some collinear
infrared (IR) cutoff. In \cite{Kovchegov:2007vf} it was demonstrated
that, for the inclusive production cross section of gluons and
collinear GG or $q \bar q$ pairs with the invariant mass less than
$\Lambda_\text{coll}$, at least one factor of the coupling is $\as
(\Lambda_\text{coll}^2)$. While it may seem tempting to use this
result and replace $\as$ in \eq{ktfact} by $\as
(\Lambda_\text{coll}^2)$ constructing a guess for the final answer, it
is not even clear a priori that \eq{ktfact} retains its
fixed-coupling, $k_T$-factorized form after the running coupling
corrections are included. Moreover, there are factors of fixed
couplings in Eqs.\ (\ref{ktglueA}) and (\ref{ktgluep}) for which one
also has to specify the momentum scale.  Therefore the problem of
setting the scale of the running couplings for the gluon production
cross section needs to be addressed by a detailed calculation, which
we will perform here.\footnote{A related problem of setting the scale
  of the running coupling in the formula for the heavy quark pair
  production from \cite{Catani:1990eg,Collins:1991ty} was studied in
  \cite{Levin:1991ya}. However, in \cite{Levin:1991ya} the large
  masses of the heavy quarks lead to a kinematic regime different from
  the one considered here.}

In this paper we will tackle this problem by calculating the running
coupling corrections to the lowest order ($O (\as^3)$) gluon
production cross section
\cite{Kuraev:1976ge,Kovchegov:1997ke,Kovner:1995ts,Kovner:1995ja}. Our
lowest-order result will allow us to conjecture how \eq{ktfact} would
be modified due to running coupling corrections. We will calculate the
running coupling corrections using the scale-setting prescription due
to Brodsky, Lepage, and Mackenzie (BLM) \cite{BLM}. This prescription
assumes that the scale of the coupling set by the introduction of
resummed single-quark-loop corrections is the scale of the coupling
set by the full calculation including both quark and gluon loop
corrections to the propagators and vertices.  After finding the $N_f$
corrections due to resummed quark loops, one `completes the beta
function' by replacing the $N_f$ in the corrections with $-6 \pi
\beta_2$. (Since we will be resumming powers of $\as \, N_f$ the
calculation will not employ the large-$N_c$ limit.)

The paper is structured as follows. We begin in \Sec{fixed} by briefly
reviewing the fixed-coupling lowest-order calculation of the high
energy inclusive gluon production cross section in quark-quark
scattering. We then include running coupling corrections in \Sec{run}.
We begin with the easier bremsstrahlung diagrams in \Sec{brem}, and
then move on to the case of diagrams with the triple-gluon vertex in
\Sec{3G}. The final result for the gluon production cross section is
given by Eqs.\ (\ref{rc_incl}) and (\ref{Qscale}) in \Sec{ans}. We
conclude by conjecturing the running coupling generalization of
\eq{ktfact} given by \eq{rc_fact} in \Sec{conj} and by discussing the
implications of our result on the CGC predictions for particle
multiplicity $dN/d\eta$ in heavy ion collisions as a function of the
centrality of the collision in \Sec{dNde}.


\section{Brief Review of the Fixed-Coupling Calculation}
\label{fixed}

Consider the leading-order contributions to the production of a gluon
in a high-energy quark-quark scattering in standard Feynman
perturbation theory, which we will use throughout the paper.  We set
up the problem in the experimentally relevant momentum regime such
that one incoming quark has very large ``$+$'' momentum
$p_1^\mu=(p_1^+,0,\wv{0})$, where we use light-cone coordinates
$(+,-,\wv{\perp})$ throughout the paper with the normalization $v^\pm
= (v^0 \pm v^3)/\sqrt{2}$ for a 4-vector $v^\mu$, and the other
incoming quark has very large ``$-$'' momentum
$p_2^\mu=(0,p_2^-,\wv{0})$.  We take $p_1^+$ and $p_2^-$ to have the
same order of magnitude, and $p_1^+$ and $p_2^-$ are large compared to
any other momentum scale in the problem.  This is the eikonal
approximation, and, in particular, $p_1^+$ and $p_2^-$ are large
compared to the transverse momentum of the final state particles.
Throughout the paper we will work in the light-cone gauge, $\eta\cdot
A = A^+ = 0$, with $\eta^\mu = (0, 1, {\bm 0})$.  In this case the
three dominant leading order Feynman diagrams are shown in
\fig{lo_gluon}, which we will refer to as the triple-gluon vertex (A)
and bremsstrahlung (B and C) diagrams.  Note that the two
bremsstrahlung diagrams that can be drawn with an emitted gluon
connecting to the lower quark line are suppressed due to eikonality in
this $A^+=0$ light-cone gauge: an easy way to see this is to notice
that at high energy the gluon emission from the lower quark line
should come in to leading order with a factor of $\gamma^-$ in the
quark-gluon vertex, which would be multiplied by the gluon
polarization component $\epsilon^+_\lambda$, which, in turn, is zero
in $A^+ = 0$ light-cone gauge.

\FIGURE{\includegraphics[width=\textwidth]{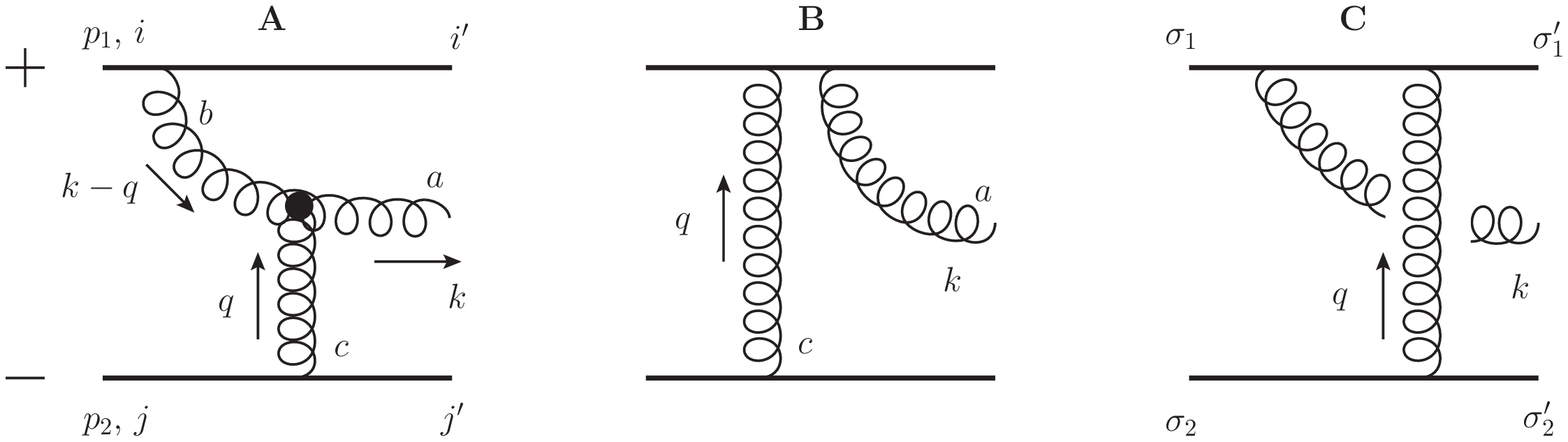}
\vspace{-.3in}
  \caption{Diagrams contributing to the lowest-order gluon production in 
    quark-quark scattering at high energy in the $A^+ =0$ light-cone
    gauge. The blob in A denotes a triple-gluon vertex while the
    outgoing gluon line in C is visualized as disconnected for
    clarity (there are only two gluon lines in C). The initial and
    final quark momenta and fundamental color indices are indicated on
    diagram A, the adjoint color indices are shown in diagrams A
    and B, and the initial and final quark helicities are indicated
    on diagram C.}
  \label{lo_gluon}
}

In a process with three final state on-shell particles, there are five
unconstrained momentum components in the cross section; we choose to
take these components to be the components of the transverse momentum
transferred from the lower quark line, $\wv{q}$, and the components of
the transverse momentum and the plus momentum of the emitted gluon,
$\wv{k}$ and $k^+$, respectively.

On-shellness of the final particles in \fig{lo_gluon} in particular
gives for the top quark line $0 = (p_1 - k+q)^2 \approx - 2 \, p_1^+
\, (k^- - q^-)$, such that
\begin{align}\label{q-}
  q^- \, = \, k^- \, = \, \frac{\wv{k}^2}{2k^+}
\end{align}
to eikonal accuracy (i.e.\ to leading order in inverse powers of the
large momenta, $p_1^+$ and $p_2^-$). Similarly imposing the on-shell
condition on the outgoing quark line at the bottom of \fig{lo_gluon}
yields
\begin{align}\label{q+}
  q^+ \, = \, 0
\end{align}
with the same accuracy. 

The light-cone gauge propagator without loop corrections is
\begin{align}
\label{gaugepropwithout}
  \frac{-i}{q^2 + i \, \epsilon} \, D_{\mu\nu} (q) \, = \,
  \frac{-i}{q^2 + i \, \epsilon} \, \left[ g_{\mu\nu} - \frac{\eta_\mu
      \, q_\nu + \eta_\nu \, q_\mu}{\eta \cdot q} \right],
\end{align}
and the outgoing gluon polarization vector $\epsilon_\lambda$ obeys
$\eta\cdot\epsilon_\lambda=\epsilon_\lambda^+=0$ and
$k\cdot\epsilon_\lambda=0$ for the two polarizations $\lambda = 1$, 2;
one may then immediately set
\begin{align}
\label{pol}
\epsilon_\lambda=\epsilon_\lambda^*=\left(0, 
\frac{\wv{k} \cdot \wv{\epsilon}}{k^+},\wv{\epsilon}\right).
\end{align}

Exploiting the eikonality of the process, we can approximate
\begin{align}
  \bar{u}_{\sigma_2'}(p_2-q) \, \gamma^\mu \, u_{\sigma_2}(p_2) \simeq
  2 \, p_2^\mu \, \delta_{\sigma_2',\sigma_2},
\end{align}
where $\sigma_2'$ and $\sigma_2$ are the helicities of the outgoing
and incoming lower line quarks, respectively \cite{Forshaw:1997dc}.  A similar
expression holds for the top quark line.  The anti-commutation
relations of the gamma matrices along with the Dirac equation yields
the useful relation \cite{Peskin:1995ev}
\begin{align}
  \bar{u}(p) \, \slashed{\epsilon} \, \slashed{p} = 2 \,
  p\cdot\epsilon \, \bar{u}(p);\qquad \slashed{p} \,
  \slashed{\epsilon} \, u(p) = 2 \, p\cdot\epsilon \, u(p).
\end{align}
Finally, we will also find the following type of cancellation useful
\begin{align}
  \bar{u}(p_2) \, \slashed{q} \, u(p_2) =
  \bar{u}(p_2-q)\big(-(\slashed{p}_2 - \slashed{q}) +
  \slashed{p}_2\big)u(p_2) = 0,
\end{align}
by the Dirac equation.

With the above notation and machinery one may almost immediate write
down the leading order in eikonality result for the bremsstrahlung
diagrams,
\begin{align}
  i\mathcal{M}_B & \simeq \phantom{-} \,
  \frac{8ig^3\,p_1^+p_2^-}{\wv{q}^2\,\wv{k}^2} \,
  \wv{\epsilon}_{\lambda} \cdot \wv{k} \, \big(t^a t^c\big)_{i',i}
  \big(t^c\big)_{j',j} \, \delta_{\sigma_1',\sigma_1} \, \delta_{\sigma_2',\sigma_2} \\
  i\mathcal{M}_C & \simeq - \,
  \frac{8ig^3\,p_1^+p_2^-}{\wv{q}^2\,\wv{k}^2} \,
  \wv{\epsilon}_{\lambda} \cdot \wv{k} \, \big(t^c t^a\big)_{i',i}
  \big(t^c\big)_{j',j} \, \delta_{\sigma_1',\sigma_1} \,
  \delta_{\sigma_2',\sigma_2},
\end{align}
where repeated indices are summed over.  A slightly more involved
calculation yields
\begin{align}
  i\mathcal{M}_A \simeq - \,
  \frac{8g^3\,p_1^+p_2^-}{(\wv{k}-\wv{q})^2\,\wv{q}^2} \,
  \wv{\epsilon}_{\lambda} \cdot (\wv{k}-\wv{q}) \, f^{abc} \,
  \big(t^b\big)_{i',i} \big(t^c\big)_{j',j} \,
  \delta_{\sigma_1',\sigma_1} \, \delta_{\sigma_2',\sigma_2}
\end{align}
for the triple-gluon vertex diagram to leading order in eikonality.

Summing the results for $\mathcal{M}_A$, $\mathcal{M}_B$, and
$\mathcal{M}_C$ and using the commutation relations for the color
matrices one finds that the matrix element for gluon production to
leading order in fixed coupling $\alpha_s$ and eikonality is
\begin{align}
  i(\mathcal{M}_A + \mathcal{M}_B+\mathcal{M}_C) =
  \frac{8g^3\,p_1^+p_2^-}{\wv{q}^2} \, \wv{\epsilon}_{\lambda} \cdot
  \left(\frac{\wv{k}}{\wv{k}^2}-\frac{\wv{k}-\wv{q}}{(\wv{k}-\wv{q})^2}\right)
  \,f^{abc} \, \big(t^b\big)_{i',i} \big(t^c\big)_{j',j}
  \delta_{\sigma_1',\sigma_1}\delta_{\sigma_2',\sigma_2}
\end{align}
After summing over final states, averaging over initial states, and
including the kinematic factors one arrives at the following
expression for the gluon production cross section
\cite{Kuraev:1976ge,Kovchegov:1997ke,Kovner:1995ts,Kovner:1995ja}:
\begin{align}\label{lip1}
  \frac{d \sigma}{d^2 k_T \, dy} \, = \, \frac{2 \, \as^3 \,
    C_F}{\pi^2} \, \int \frac{d^2 q}{({\bm q}^2)^2} \, \sum_{\lambda}
  \, \bigg| \frac{{\bm \epsilon}_\lambda \cdot ({\bm k} - {\bm
      q})}{({\bm k} - {\bm q})^2} - \frac{{\bm \epsilon}_\lambda \cdot
    {\bm k}}{{\bm k}^2} \bigg|^2.
\end{align}

Summing over gluon polarizations and opening the brackets in
\eq{lip1} yields
\begin{align}\label{lip2}
  \frac{d \sigma}{d^2 k_T \, dy} \, = \, \frac{2 \, \as^3 \,
    C_F}{\pi^2} \, \frac{1}{{\bm k}^2} \, \int \, \frac{d^2 q}{{\bm
      q}^2 \, ({\bm k} - {\bm q})^2}. 
\end{align}
Our goal now is to set the scales for the three couplings in
\eq{lip2}. Before we do that let us note that the lowest-order gluon
distribution of a quark is (see e.g.
\cite{Jalilian-Marian:1997xn,Kharzeev:2003wz})
\begin{align}\label{phi_lo}
  \phi ({\bm k}, y) \, = \, \frac{\as \, C_F}{\pi} \, \frac{1}{{\bm k}^2}.
\end{align}
With the help of \eq{phi_lo} we can see that \eq{lip2} reduces exactly
to \eq{ktfact} for quark-quark scattering.


\section{Running Coupling Corrections}
\label{run}

To include running coupling corrections we will follow the BLM
scale-setting procedure \cite{BLM}. We will first resum the
contribution of all quark bubble corrections giving powers of $\amu \,
N_f$, with $N_f$ the number of quark flavors and $\amu$ the physical
coupling at some arbitrary renormalization scale $\mu$. We will then
complete $N_f$ to the full beta-function by replacing
\begin{align}\label{repl}
N_f \rightarrow - 6 \, \pi \, \beta_2
\end{align} 
in the obtained expression. Here
\begin{align}\label{beta}
\beta_2 = \frac{11 N_c - 2 N_f}{12 \, \pi}  
\end{align} 
is the one-loop QCD beta-function. After this, the powers of $\amu \,
\beta_2$ should combine into physical running couplings 
\begin{align}\label{as_geom}
  \as (Q^2) \, = \, \frac{\amu}{1 + \amu \, \beta_2 \, \ln
    \frac{Q^2}{\mu^2}}
\end{align}
at various momentum scales $Q$ which would follow from this
calculation. Throughout this paper we will use the $\overline {\text
  MS}$ renormalization scheme.

Below we will begin by including running coupling corrections into the
bremsstrahlung diagram B from \fig{lo_gluon} assuming that only a
gluon can be produced in the final state. By doing so we will not be
able to specify the scale for one of the coupling constants in the
diagram, obtaining a contribution to the cross section that depends on
an arbitrary renormalization scale $\mu$. Following
\cite{Kovchegov:2007vf} we will rectify the problem by redefining the
gluon production cross-section to include production of collinear
gluon--gluon and quark--anti-quark pairs with the invariant mass lower
than some collinear IR cutoff $\Lambda_\text{coll}^2$. We will see
that this new observable will be completely $\mu$-independent and
expressible in terms of the running coupling constants.


\subsection{Bremsstrahlung Diagrams}
\label{brem}

We begin by considering bremsstrahlung diagrams B and C in
\fig{lo_gluon}. It is easier to see how running coupling corrections
are organized if we consider the production cross section, which is
given by the square of the sum of the diagrams in \fig{lo_gluon}.
Consider the square of the diagram B shown in the left panel of
\fig{Bsq}. Let us include all the quark loop corrections to it which
bring in powers of $\amu \, N_f$.  The non-vanishing quark loop
corrections to diagram B from \fig{lo_gluon} squared are shown in the
right panel of \fig{Bsq}, where we imply that quark loops need to be
resummed to all orders on the propagators where we inserted several
loops. The corrections are limited to quark bubbles on the propagator
of the gluon line carrying momentum $q$ in the amplitude and in the
complex conjugate amplitude. The (massless) quark loop corrections on
the outgoing gluon line are zero in dimensional regularization, since
the produced gluon is on mass shell, $k^2 =0$ \cite{Sterman:1994ce}.
Throughout the paper we assume that the quarks in the loops are
massless: quark mass is not needed to fix the scale of the coupling
constants.

\FIGURE{\includegraphics[width=12.5cm]{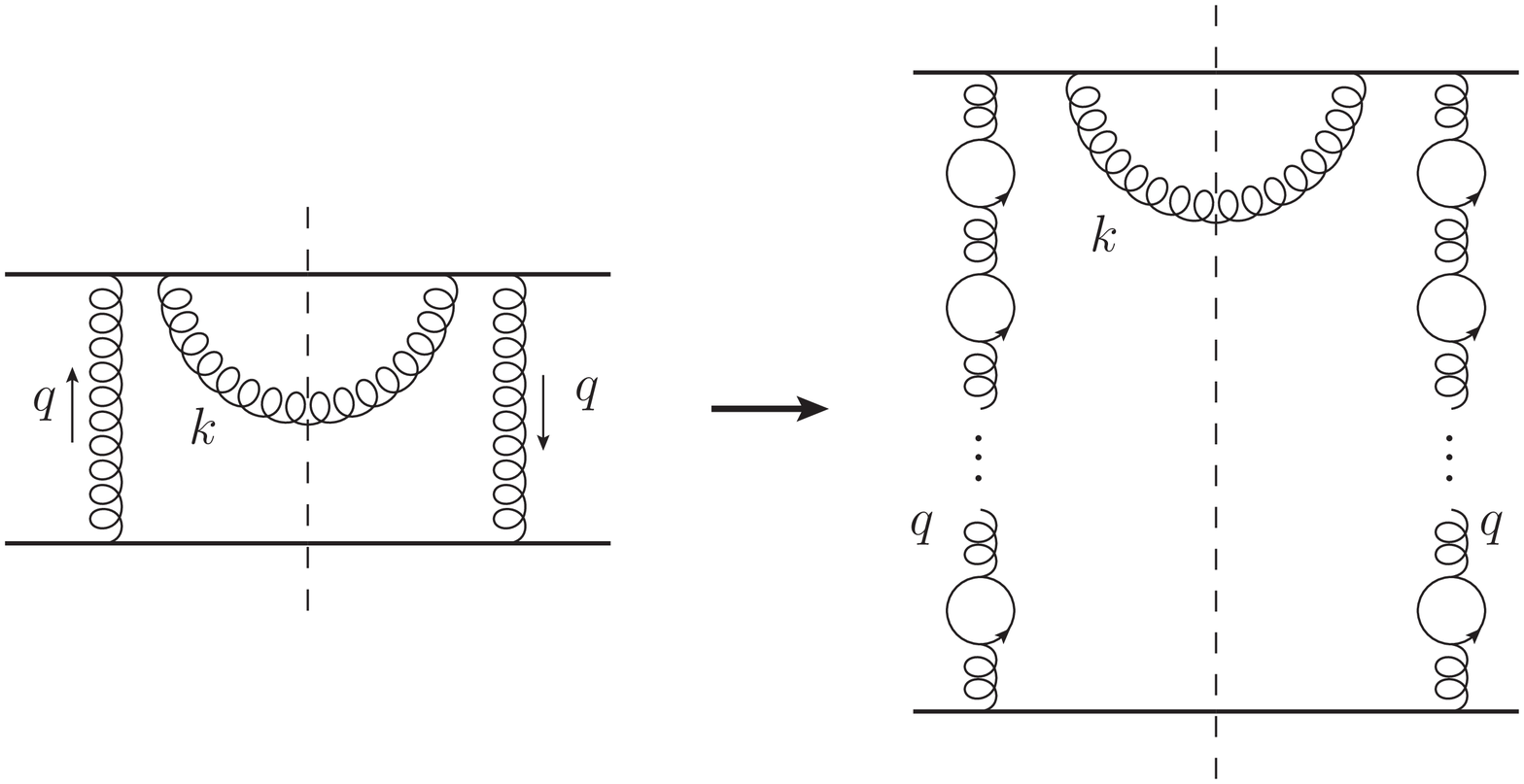}
  \caption{Inclusion of quark loop corrections to diagram B from \fig{lo_gluon} 
    squared. Vertical dashed line denotes the final state cut.}
  \label{Bsq}
}

The result of the resummation of quark bubble corrections to a gluon
propagator is well-known (see e.g.
\cite{Peskin:1995ev,Sterman:1994ce}), and this result is particularly
simple for the gluon propagator in the light-cone gauge that we are
working in. With the form of the light-cone gauge propagator without
loop corrections, \eq{gaugepropwithout}, one can straightforwardly
show that, if the contribution from each quark loop is written as
\begin{align}
  [q^2 \, g_{\alpha \beta} - q_\alpha \, q_\beta] \, i \, \Pi (q^2)
\end{align}
then the resummed gluon propagator is
\begin{align}
  \label{eq:resumglueprop}
  \frac{-i}{(q^2 + i \, \epsilon) \, [1 - \Pi (q^2)]} \, D_{\mu\nu}
  (q).
\end{align}
One can see that the tensor structure of the bare propagator,
\eq{gaugepropwithout}, is not modified by the loop corrections,
\eq{eq:resumglueprop}. An explicit calculation of $\Pi (q^2)$ to
leading order due to the quark loop and the $N_f$ part of the gluon
propagator counterterm in the $\overline{\mbox{MS}}$ renormalization
scheme yields \cite{Sterman:1994ce}
\begin{align}\label{dress_prop}
  \frac{-i}{(q^2 + i \, \epsilon) \, \left[1 - \frac{\amu \, N_f}{6 \,
        \pi} \, \ln \frac{-q^2-i\epsilon}{\mu^2} \right]} \,
  D_{\mu\nu} (q)
\end{align}
where 
\begin{align}\label{mmsbar}
  \mu^2 \, = \, \mu^2_{\overline {\text{MS}}} \ e^{5/3}.
\end{align}
We introduce a graphical shorthand for this resummed gluon propagator
in \fig{resumglueprop}.  In the same figure we also define for later
convenience the cut of this resummed gluon propagator.

\FIGURE{\includegraphics[width=16cm]{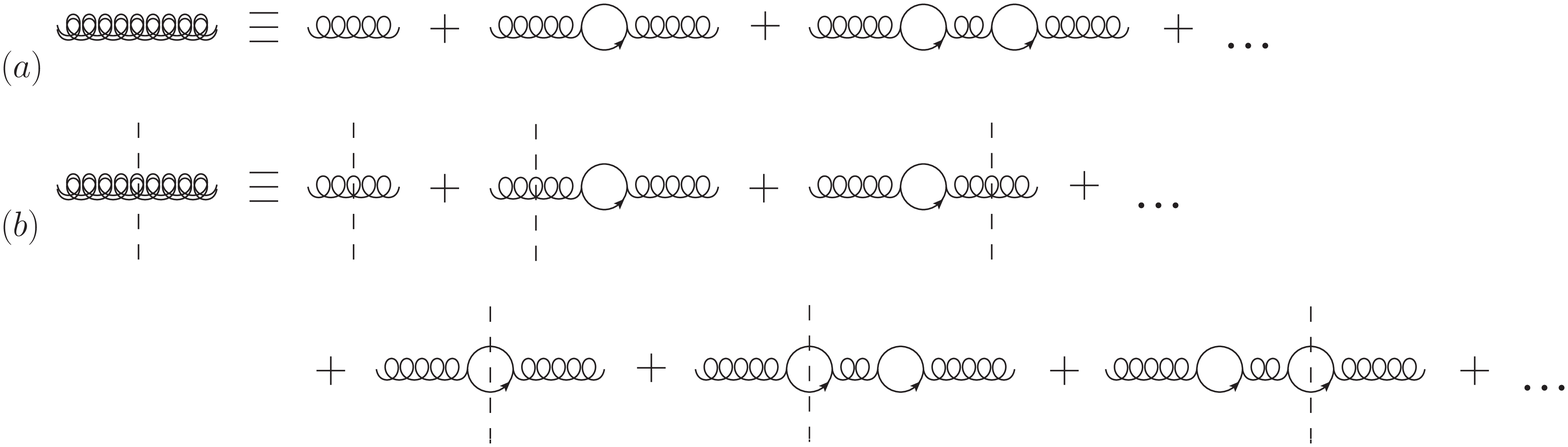}
  \caption{(a) Graphical shorthand for the resummed gluon propagator, 
    the bare gluon propagator dressed by an infinite sum of one loop
    quark loops.  (b) Definition of the cut of the resummed gluon
    propagator, defined here for later convenience.  Even though the
    cuts through gluon propagators on lines with massless quark loops
    are zero in dimensional regularization \cite{Sterman:1994ce} we
    keep them explicit here for generality.}
  \label{resumglueprop}
}

Employing \eq{dress_prop} in evaluating the right panel of \fig{Bsq},
completing $N_f$ to the full beta function using \eq{repl}, and using
the definition of the running coupling, \eq{as_geom}, we see that the
fixed couplings of diagram B in \fig{lo_gluon} squared are modified by
loop corrections to
\begin{align}\label{2out3}
  \amu^3 \, \rightarrow \, \frac{\amu^3}{\left[ 1 + \amu \, \beta_2 \,
      \ln \frac{{\bm q}^2 \, e^{-5/3}}{\msbar^2} \right]^2} \, = \,
  \amu \, \as^2 \left( {\bm q}^2 \, e^{-5/3} \right).
\end{align}

We clearly have a problem, since one of the factors of the coupling is
still taken at some arbitrary renormalization scale $\mu$.  This
problem cannot be mitigated simply by the inclusion of the corrections
due to diagram A, as the above result is all one would obtain in QED
for photon production (bremsstrahlung).  It seems that the scale of
the coupling is not fixed uniquely for the diagram in \fig{Bsq}. As
was shown in \cite{Kovchegov:2007vf} the problem can be remedied if
one redefines the observable: on top of the gluon production we should
allow for the production of collinear gluon-gluon (GG) and
quark--anti-quark ($q \bar q$) pairs. This means that instead of
calculating the diagram on the right-hand-side of \fig{Bsq}, one has
to calculate the graph pictured on the left of \fig{Bloops}.  As is
shown in \fig{resumglueprop}, the cut through the resummed gluon
propagator in \fig{Bloops} includes the contribution from the square
of diagram B in \fig{lo_gluon} and also cuts through both gluon
propagators with the gluon line dressed with quark loops and cuts
through the quark loops themselves.  Even though the contribution from
the cuts through gluon propagators with quark loop dressings are zero,
as mentioned above, we will keep them explicit for the sake of
generality (it will turn out that they are exactly canceled by a
contribution from part of the cuts through the quark loops). Because
of the redefined observable we are calculating, in addition to the
gluon production resulting from the cuts through gluon propagators we
now have to include the collinear $q \bar q$ pair production from cuts
through the quark loops.  Since we are calculating quark loop
corrections to sum up powers of $\amu \, N_f$, we do not need to
explicitly include collinear gluon pair production in the calculation:
it will be included when we complete $N_f$ to the full beta-function.
(Note that we are only interested in the terms with collinear
singularities that contribute to the running of the coupling.  Also
note that -- by the BLM prescription -- by completing to the full
beta-function we are including contributions from, e.g., the cut gluon
loop, and therefore the running coupling contribution coming from
collinear gluons.)

\FIGURE{\includegraphics[width=14cm]{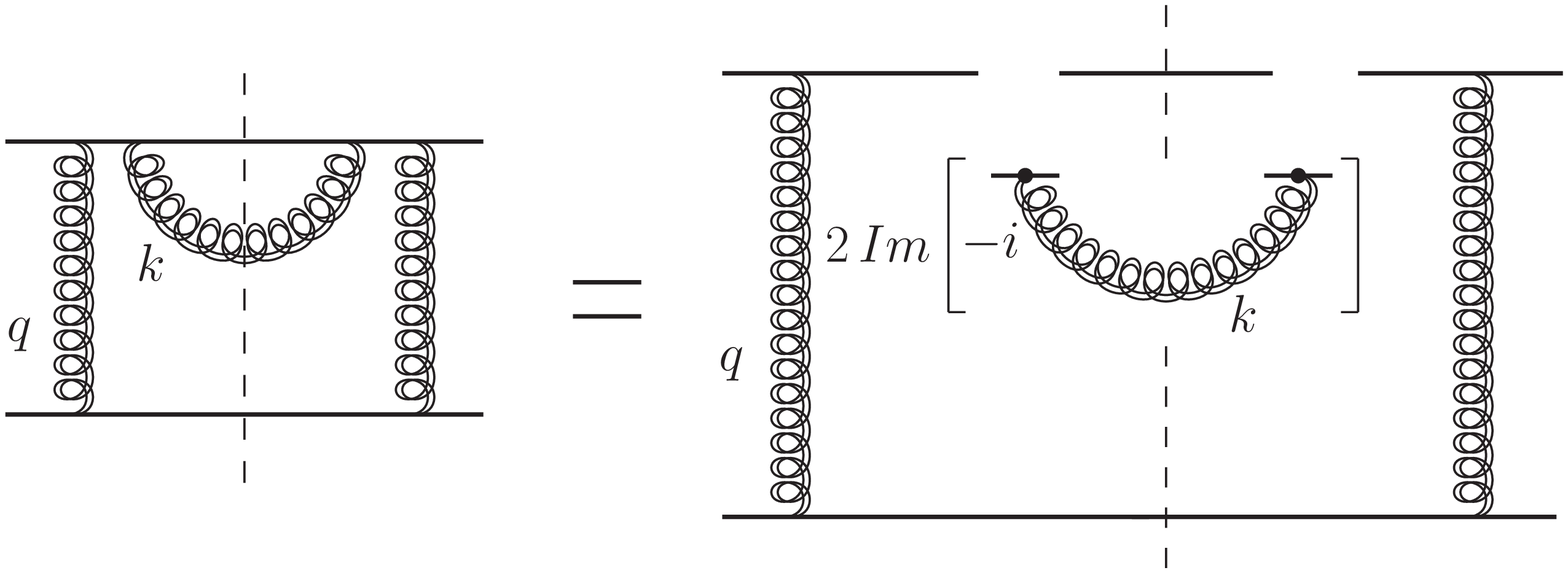}
  \caption{Quark loop corrections to the diagram B in \fig{lo_gluon} 
    for the calculation of the observable where one tags on both the
    gluons and collinear GG and $q \bar q$ pairs. The left side of the equation
    indicates the cut diagrams to be summed. The right side of the equation explains a
    simple way of performing the summation (see text). The $k$-line gluon propagator in the
    lower panel connects to the upper quark line as indicated by
    quark-gluon vertices at its ends, though this part of the diagram
    is separated from the rest for illustrative purposes.}
  \label{Bloops}
}

To sum the diagrams represented by the diagram on the left side of
\fig{Bloops} we note that the sum of the cut diagrams can be written
as the imaginary part of a dressed gluon propagator, as shown by the
diagram on the right side of \fig{Bloops}. The contribution of the
dressed propagator is calculated using \eq{dress_prop}. Including the
adjacent quark-gluon vertices one gets
\begin{multline}\label{dr_prop1}
  \int \frac{d^4 k}{(2 \, \pi)^4} \, 2 \, \text{Im} \, \left[ i \,
    \frac{-i \, \amu}{(k^2 + i \, \epsilon) \, \left[1 + \amu \,
        \beta_2 \, \ln \frac{-k^2-i\epsilon}{\mu^2} \right]} \, D_{\mu\nu} (k)
  \right] \, = \\[5pt] \int \frac{d^4 k}{(2 \, \pi)^4} \, 2 \, \text{Im} \,
  \left[ \frac{\as \left( -k^2 \, e^{-5/3} -i\epsilon\right)}{k^2 + i \,
      \epsilon} \, D_{\mu\nu} (k) \right],
\end{multline}
where, according to the optical theorem, we took the imaginary part of
the ``forward amplitude'' and multiplied it by $2$. The factor of $i$
results from the $-i$ needed to convert $i \, M$ into the amplitude
$M$ and from $i^2 = -1$ coming from the quark-gluon vertices. In the
eikonal approximation used here the indices of the gluon propagator
coupling to the upper quark line carrying a large ``$+$'' component of
momentum are $\mu=\nu=+$. However, we will keep the indices $\mu, \nu$
unspecified, as this will make our discussion more general. In
arriving at \eq{dr_prop1} we have also completed $N_f$ to the full
beta function, which allowed us to replace the geometric series by a
factor of the running coupling constant. \eq{dr_prop1} is illustrated
on the right side of \fig{Bloops}: it is a short hand way of taking
the imaginary part of the whole diagram, keeping only the cuts going
through the dressed gluon propagator with momentum $k$.

\eq{dr_prop1} is still somewhat incomplete, since we need to specify
what we mean by the collinear $q \bar q$ pairs. To remedy this problem
we write using the spectral representation (see
\cite{Dokshitzer:1978hw,Dokshitzer:1993pf} for a similar approach)
\begin{align}\label{spectr}
  \int \frac{d^4 k}{(2 \, \pi)^4} \, = \, \int\limits_{-\infty}^\infty
  dk^2 \, \int \frac{d k^+ \, d^2 k_\perp}{2 \, k^+ \, (2 \, \pi)^4}
  \, \Rightarrow \, \int\limits_0^{\Lambda_\text{coll}^2} dk^2 \, \int
  \frac{d k^+ \, d^2 k_\perp}{2 \, k^+ \, (2 \, \pi)^4},
\end{align}
where $k^2 = k_\mu \, k^\mu$ is the virtuality of the gluon line.  The
last step of \eq{spectr} completes the definition of our observable.
By limiting the virtuality of the gluon line between $0$ and some
small momentum scale $\Lambda_\text{coll}^2 >0$, we keep the diagrams
with the gluon line in the final state (top line of
\fig{resumglueprop} (b)) and limit the invariant mass of the produced
$q\bar q$ pair (bottom line of \fig{resumglueprop} (b)) to be smaller
than $\Lambda_\text{coll}^2$, which ensures the collinearity of the
pair. (One can show that a small-virtuality $q\bar q$ pair with both
quarks on-shell and massless necessarily has the two quarks collinear
with each other.)  $\Lambda_\text{coll}$ plays the role of the minimum
momentum scale to be resolved by the ``detector''.  Below we will
assume that $\Lambda_\text{coll}$ is much smaller than all the
momentum scales involved in the problem, but is much larger than
$\Lambda_{QCD}$, such that perturbation theory remains applicable.
Resummation of higher-order collinear emissions would likely lead to a
Sudakov form-factor \cite{Sudakov:1954sw}, reducing the dependence of
the cross section on $\Lambda_\text{coll}$.  (A detailed discussion of
the role of $\Lambda_\text{coll}$ can be found in
\cite{Kovchegov:2007vf}, where it is argued that $\Lambda_\text{coll}$
plays the role of the factorization scale for the fragmentation
functions.)

Using \eq{spectr} in \eq{dr_prop1} yields
\begin{align}\label{dr_prop2}
  \Delta_{\mu\nu} \, \equiv \, \int\limits_0^{\Lambda_\text{coll}^2}
  dk^2 \, \int \frac{d k^+ \, d^2 k_\perp}{k^+ \, (2 \, \pi)^4} \,
  \text{Im} \, \left[ \frac{\as \left( -k^2 \, e^{-5/3}
        -i\epsilon\right)}{k^2 + i \, \epsilon} \, D_{\mu\nu} (k)
  \right].
\end{align}
Since we assume that $\Lambda_\text{coll}$ is the smallest
perturbative momentum scale in the problem, we are interested only in
the terms which do not vanish in the $\Lambda_\text{coll} \rightarrow
0$ limit. Since the numerator of the gluon propagator $D_{\mu\nu} (k)$
is completely real and does not have any singularities at $k^2 =0$
(corresponding to the $\Lambda_\text{coll} \rightarrow 0$ limit), we can
move $D_{\mu\nu} (k)$ outside the integral over gluon virtuality $k^2$
and outside the $\text{Im}$ part sign. This modifies \eq{dr_prop2}
into
\begin{align}\label{dr_prop3}
  \Delta_{\mu\nu} \, \approx \, \int \frac{d k^+ \, d^2 k_\perp}{k^+
    \, (2 \, \pi)^4} \, D_{\mu\nu} (k) \,
  \int\limits_0^{\Lambda_\text{coll}^2} dk^2 \, \text{Im} \, \left[
    \frac{\as \left( -k^2 \, e^{-5/3} -i\epsilon\right)}{k^2 + i \,
      \epsilon} \right].
\end{align}
Extracting the imaginary part in \eq{dr_prop3} we obtain
\begin{align}\label{dr_prop4}
  \Delta_{\mu\nu} \, \approx \, \int \frac{d k^+ \, d^2 k_\perp}{k^+
    \, (2 \, \pi)^4} \, D_{\mu\nu} (k) \,
  \int\limits_0^{\Lambda_\text{coll}^2} dk^2 \, \left[ - \pi \, \delta
    (k^2) \, \as (0) + \frac{1}{k^2} \, \frac{\pi \, \beta_2 \,
      \amu^2}{\left( 1 + \amu \, \beta_2 \, \ln \frac{k^2}{\mu^2}
      \right)^2 + \pi^2 \, \beta_2^2 \, \amu^2} \right]
\end{align}
where for the moment we do not need to specify what we mean by $\as
(0)$. Integrating over $k^2$ in \eq{dr_prop4} yields
\begin{align}\label{dr_prop5}
  \Delta_{\mu\nu} \, & \approx \, \int \frac{d k^+ \, d^2 k_\perp}{k^+
    \, (2 \, \pi)^4} \, D_{\mu\nu} (k) \notag \\ & \times \, \left\{ -
    \pi \, \as (0) + \frac{1}{\beta_2} \, \left[ \arctan \left(
        \frac{1}{\pi \, \beta_2 \, \as \left( \Lambda_\text{coll}^2 \,
            e^{-5/3} \right)} \right) - \arctan \left( \frac{1}{\pi \,
          \beta_2 \, \as (0)} \right) \right] \right\}.
\end{align}
We see now that all the factors of $\amu$ got absorbed into running
coupling constants at different momentum scales. However, since we are
aiming to find the scale of one power of the coupling $\amu$, we do
not have control over higher powers of the coupling $\as$ which enter
\peq{dr_prop5}. Therefore we need to expand \eq{dr_prop5} to the
lowest order in $\as$. This gives
\begin{align}\label{dr_prop6}
  \Delta_{\mu\nu} \, \approx \, \int \frac{d k^+ \, d^2 k_\perp}{k^+
    \, (2 \, \pi)^4} \, D_{\mu\nu} (k) \, (- \pi) \, \as \left(
    \Lambda_\text{coll}^2 \, e^{-5/3} \right) \, = \, \as \left(
    \Lambda_\text{coll}^2 \, e^{-5/3} \right) \, \int \frac{d k^+ \,
    d^2 k_\perp}{2 \, k^+ \, (2 \, \pi)^3} \, \sum_\lambda \,
  \epsilon_\mu^\lambda (k) \, \epsilon_\nu^{\lambda \, *} (k)
\end{align}
with the polarization vector $\epsilon_\mu^\lambda$ given by \eq{pol}
and $*$ denoting complex conjugation.  We see that in \eq{dr_prop6} we
obtained all the standard factors for the outgoing gluon line, along
with the running coupling constant $\as \left( \Lambda_\text{coll}^2
  \, e^{-5/3} \right)$. The kinematic factors and polarizations are
the same as for the fixed coupling diagram B from \fig{lo_gluon}
squared. We see that the scale of the remaining coupling constant in
\eq{2out3} has now been fixed and is given by $\Lambda_\text{coll}^2
\, e^{-5/3}$ in the $\overline {\text MS}$ renormalization
scheme.\footnote{Note that for quarks with a nonzero but small mass
  $\Lambda_\text{coll}^2$ is replaced by $4 \, m^2$ in the argument of
  $\as$ (with $m$ the quark mass). The quark mass will regulate
  collinear divergences, making it unnecessary to introduce the IR
  cutoff $\Lambda_\text{coll}^2$.} (Note that the terms with $\as (0)$
from \eq{dr_prop5} got canceled in arriving at \eq{dr_prop6}, thus
relieving us of the need to properly define this quantity.)

Combining our result in \peq{dr_prop6} with \peq{2out3} we see that
the three fixed-coupling constants in the diagram in \fig{lo_gluon}B
squared become the following running couplings:
\begin{align}\label{brem_rc}
  \amu^3 \, \rightarrow \, \as \left( \Lambda_\text{coll}^2 \,
    e^{-5/3} \right) \, \as^2 \left( {\bm q}^2 \, e^{-5/3} \right).
\end{align}
The above analysis can be repeated for the square of the diagram in
\fig{lo_gluon}C and for the cross term between diagrams B and C in
\fig{lo_gluon}, in each case leading to the same answer given by
\eq{brem_rc}. We therefore conclude that \eq{brem_rc} gives us the
scales of the running couplings for the ``bremsstrahlung diagrams'' B
and C in \fig{lo_gluon} taken by themselves, without the diagram A. We
now will analyze the running coupling corrections to diagram A of
\fig{lo_gluon}.


\subsection{Triple-Gluon Vertex Diagrams}
\label{3G}

Quark loop corrections to the triple-gluon vertex diagram from
\fig{lo_gluon}A squared are shown in \fig{3Gloops}. We now have to
``dress'' both the $q$ and $k-q$ gluon lines with quark bubbles. The
outgoing gluon line also needs to be dressed like it was done in
\fig{Bloops} for bremsstrahlung diagrams: again the cut can go either
through the gluon line or through a quark loop. On top of that the
triple gluon vertex may receive a quark bubble correction as well.
Note that in principle the cut can go through the quark loop
correction to the triple-gluon vertex. However, it is easy to show
that that contribution does not have a singularity when the gluon
virtuality is zero, $k^2 = 0$. Integration of the contribution from
such a cut over $k^2$ from $0$ to $\Lambda_\text{coll}^2$ would thus
yield an expression proportional to $\Lambda_\text{coll}^2$, which
vanishes in the $\Lambda_\text{coll} \rightarrow 0$ limit. Since, just
like for bremsstrahlung diagrams, we are interested in the
contributions which do not vanish in the $\Lambda_\text{coll}
\rightarrow 0$ limit, the cuts through the quark bubble at the triple
gluon vertex can therefore be neglected. We see that for the outgoing
gluon propagator we have to sum over the same cuts as was done in
\Sec{brem} for the bremsstrahlung diagrams, as pictured in
\fig{Bloops}. Since our derivation of the result \peq{brem_rc} of such
a summation was valid for any values of the propagator indices $\mu$
and $\nu$, the conclusion applies to the diagrams in \fig{3Gloops} as
well: the ``dressing'' of the outgoing gluon propagator along with the
adjacent vertices gives us a factor of $\as \left(
  \Lambda_\text{coll}^2 \, e^{-5/3} \right)$.

\FIGURE{\includegraphics[width=14cm]{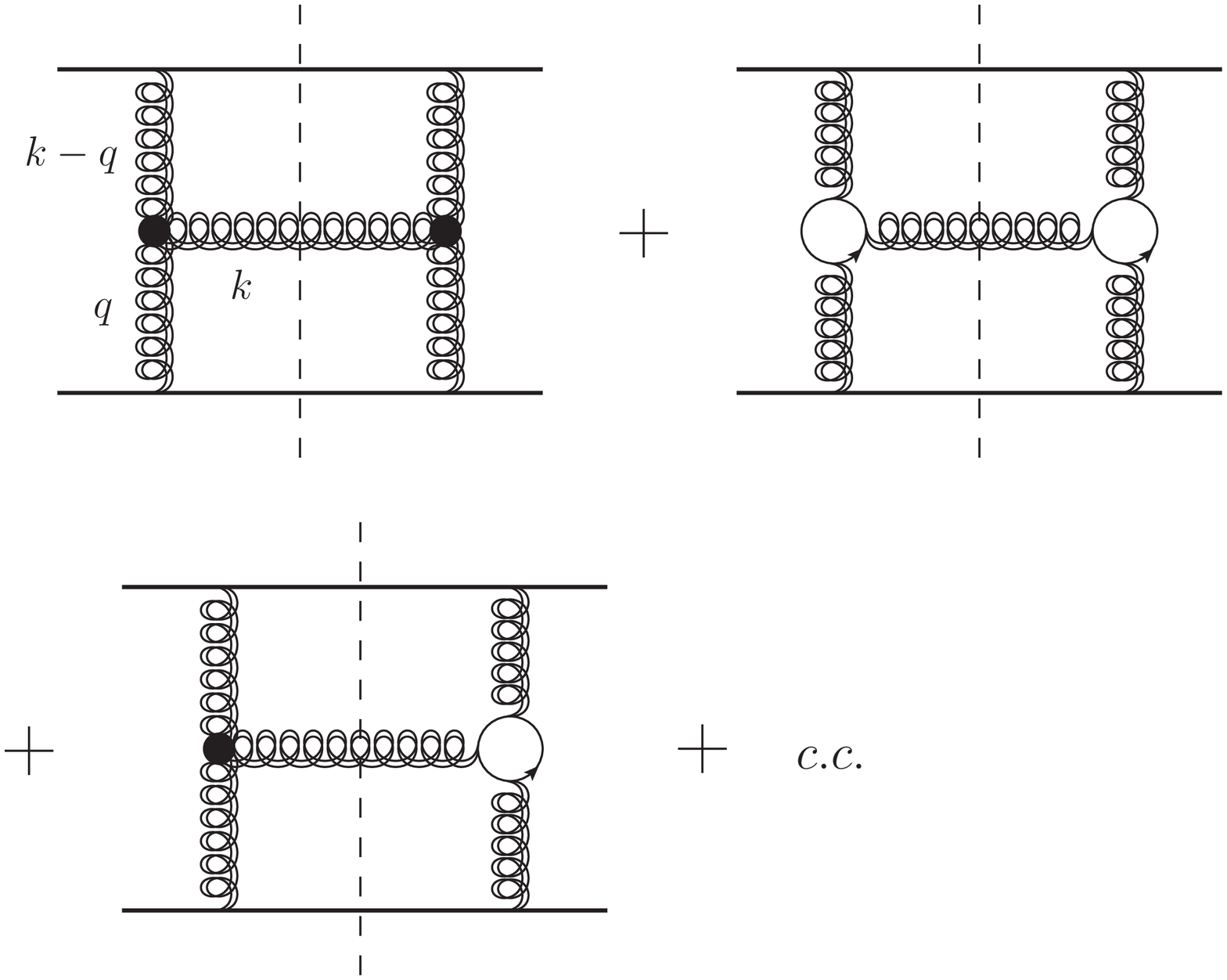}
  \caption{Quark loop corrections to the square of diagram A from 
    \fig{lo_gluon}.  The dot denotes the usual QCD triple-gluon
    vertex. }
  \label{3Gloops}
}

Concentrating on the top left diagram in \fig{3Gloops} and using
\peq{dress_prop} along with \peq{repl} we see that quark loops lead to
the following modification of the factors of coupling in it:
\begin{align}\label{overall}
  \amu^3 \, & \rightarrow \, \as \left( \Lambda_\text{coll}^2 \,
    e^{-5/3} \right) \, \frac{\amu^2}{\left[ 1 + \amu \, \beta_2 \,
      \ln \frac{{\bm q}^2 \, e^{-5/3}}{\msbar^2} \right]^2 \, \left[ 1
      + \amu \, \beta_2 \, \ln \frac{({\bm k} - {\bm q})^2 \,
        e^{-5/3}}{\msbar^2} \right]^2} \notag \\ \, & = \, \frac{\as
    \left( \Lambda_\text{coll}^2 \, e^{-5/3} \right) \, \as^2 \left(
      {\bm q}^2 \, e^{-5/3} \right) \, \as^2 \left( ({\bm k} - {\bm
        q})^2 \, e^{-5/3} \right)}{\amu^2}.
\end{align}
Indeed, since we have analyzed only one diagram we have not fixed the
scales of all the coupling constants. However, the expression in
\peq{overall} is an overall factor, present in each of the diagrams in
\fig{3Gloops}. As was noted in \cite{Kovchegov:2007vf} the effect of
the quark loop corrections to the triple gluon vertex would be to
generate factors of the type
\begin{align}\label{factrs}
  \left[ 1 + \amu \, \beta_2 \, \ln \frac{Q_A^2 \,
      e^{-5/3}}{\msbar^2}\right] \, \left[ 1 + \amu \, \beta_2 \, \ln
    \frac{Q_B^2 \, e^{-5/3}}{\msbar^2}\right]
\end{align}
which, after multiplying \eq{overall}, would turn $1/\amu^2$ in it
into $1/\left[ \as (Q_A^2 \, e^{-5/3}) \, \as (Q_B^2 \, e^{-5/3})
\right]$ with some physical momentum scales $Q_A$ and $Q_B$ to be
determined by an explicit calculation.

Let us calculate the contribution of a quark-loop vertex correction to
the diagram A from \fig{lo_gluon}. The diagram with the one-loop
correction is shown in \fig{Aloop}. Considering the loop itself (in
which we include only the quark propagators and quark-gluon vertices
along the loop) and adding to it the contribution from the loop with
the particle number in the loop flowing in the opposite direction we
find
\begin{align}\label{loop1}
  \frac{i}{2} \, g^3 \, f^{abc} \, N_f \, \int \frac{d^d l}{(2 \,
    \pi)^d} \, \frac{\Tr \left[ \gamma^\mu \, ({\sh l} + {\sh k} -
      {\sh q}) \, \gamma^\nu \, {\sh l} \, \gamma^\rho \, ({\sh l} -
      {\sh q}) \right]}{(l^2 + i \, \epsilon) \, [(l-q)^2 + i \,
    \epsilon] \, [(l+k-q)^2 + i \, \epsilon]}
\end{align}
where, in preparation for using dimensional regularization, we have
switched to $d$ dimensions in the integral. Note that, as before, to
simplify the algebra, we assume that all quarks in the loops are
massless, since this assumption does not affect the scale of the
running coupling.

Before we evaluate \peq{loop1} let us make some simplifications due to
the kinematics. The quark-gluon vertices at the top (carrying momentum
$p_1$) and at the bottom (carrying momentum $p_2$) quark lines in
\fig{Aloop} are eikonal: the dominant contribution to the vertex on
the top comes with a Dirac matrix $\gamma^+$, while the vertex at the
bottom brings in $\gamma^-$. Using these eikonal vertices along with
Eqs.\ (\ref{q-}) and (\ref{q+}) we note that the propagator of the
$k-q$ gluon line from \fig{Aloop} can be rewritten as
\begin{align}\label{prop1}
  \frac{-i}{(k-q)^2} \, D_{\alpha \nu} (k-q) \, \rightarrow \,
  \frac{-i}{k^+} \ \frac{(k-q)_\nu^\perp}{({\bm k} - {\bm q})^2}.
\end{align}
Therefore the $\nu$ index in \eq{loop1} is only transverse, as
indicated in \fig{Aloop}. The propagator of the gluon line carrying
momentum $q$ in \fig{Aloop} can be replaced by
\begin{align}\label{prop2}
  \frac{-i}{q^2} \, D_{\beta \rho} (q) \, \rightarrow \, \frac{i}{{\bm
      q}^2} \, g^{+}_{\ \rho}
\end{align}
making the $\rho$-index in \eq{loop1} equal to $+$.

All the simplifications in Eqs.\ (\ref{q-}), (\ref{q+}),
(\ref{prop1}), and (\ref{prop2}) are exactly the same as what is
usually used in arriving at the fixed-coupling expression for the
gluon production cross section \peq{lip1}. Since we are interested in
corrections to the fixed-coupling result, we do not need to keep all
the factors, since they are the same as for the running-coupling case,
too: therefore, to simplify the algebra we multiply the expression in
\eq{loop1} only by $(k-q)_\nu^\perp$ from \eq{prop1}, $g^{+}_{\ \rho}$
from \eq{prop2}, and by $\epsilon^\lambda_\mu (k)$ from \eq{pol} for
the outgoing gluon line in \fig{Aloop}. This modifies \eq{loop1} to
\begin{align}\label{loop2}
  \Gamma \, \equiv \, \frac{i}{2} \, g^3 \, f^{abc} \, N_f \,
  \epsilon^\lambda_\mu (k) \, (k-q)_\nu^\perp \, \int \frac{d^d l}{(2
    \, \pi)^d} \, \frac{\Tr \left[ \gamma^\mu \, ({\sh l} + {\sh k} -
      {\sh q}) \, \gamma^\nu_\perp \ {\sh l} \, \gamma^+ \, ({\sh l} -
      {\sh q}) \right]}{(l^2 + i \, \epsilon) \, [(l-q)^2 + i \,
    \epsilon] \, [(l+k-q)^2 + i \, \epsilon]}
\end{align}

\FIGURE{\includegraphics[width=8cm]{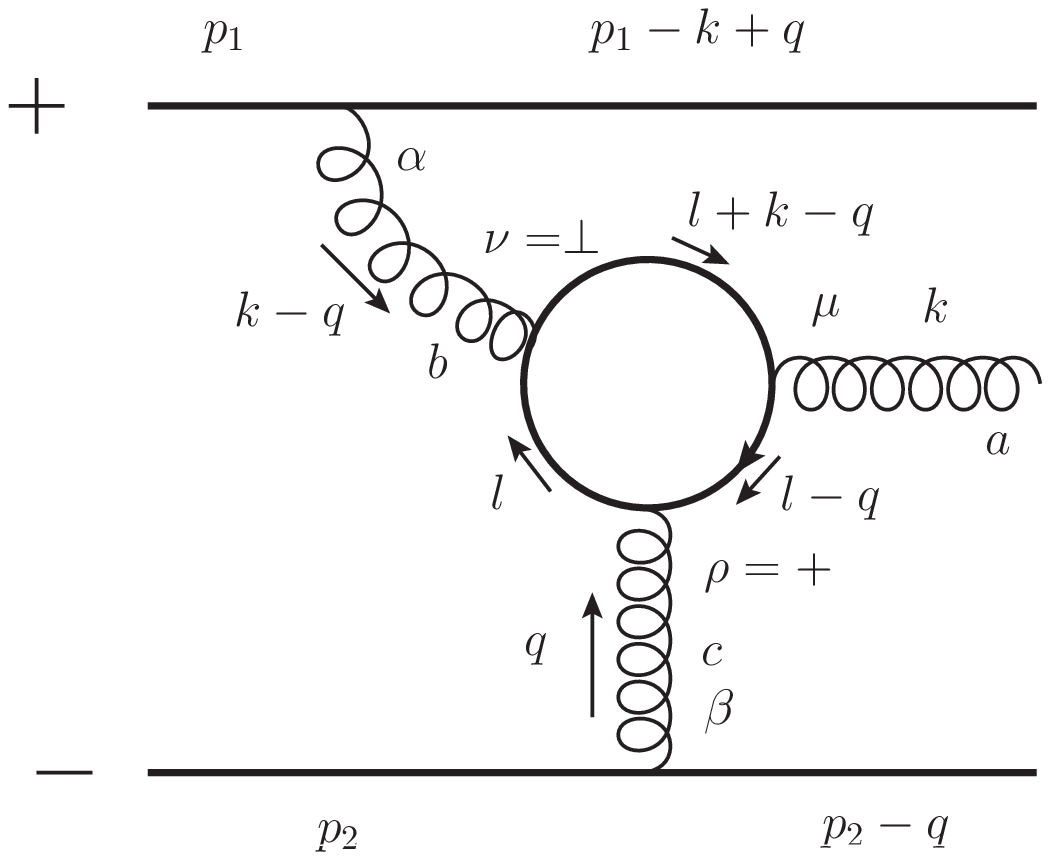}
  \caption{One quark loop correction to the triple gluon vertex in diagram A 
    from \fig{lo_gluon}. The disconnected arrows denote momentum flow
    direction, while the arrow on the quark loop denotes the particle
    number flow. The diagram with the particle number flowing in the
    opposite direction should also be included.}
  \label{Aloop}
}

The evaluation of the integral in \eq{loop2} is carried out in the
Appendix \ref{A} with the answer given in Eqs.\ \peq{Gam}, \peq{La},
\peq{Lb}, and \peq{GLO}. In arriving at \eq{Gam} we have also added
the $N_f$ part of the triple-gluon vertex counterterm to remove the
infinity in \eq{loop2}.  We note that the tensor structure of the
one-loop correction to the triple-gluon vertex is not limited to that
of the lowest-order (LO) triple-gluon vertex, which, in fact, is
well-known in the literature (see, e.g.,
\cite{Ball:1980ax,Davydychev:2001uj}).  For diagram A in
\fig{lo_gluon} the contribution of the triple-gluon vertex is given by
\eq{3Glue} and only leads to terms proportional to ${\bm
  \epsilon}_\lambda \cdot ({\bm k} - {\bm q})$ as can be seen from
\eq{3G1}. The full one-loop correction \peq{Gam} includes other tensor
structures (see \eq{Gam41}) leading to the appearance of ${\bm
  \epsilon}_\lambda \cdot {\bm k}$ terms. Of course the ultraviolet
(UV)-divergent part of the one-loop diagram does come in with the
tensor structure of the lowest-order triple-gluon vertex \peq{3Glue}.
However, in accordance with the BLM prescription, we need to collect
all $\amu \, N_f$ terms to set the scale of the coupling constants:
therefore the UV-finite term coming with ${\bm \epsilon}_\lambda \cdot
{\bm k}$ in \eq{Gam} is also included.

Since the tensor structure of the LO triple-gluon vertex is modified,
we see that the naive expectation for the quark loop correction to
simply bring in overall factors like those shown in \eq{factrs} into
the square of the diagram in \fig{3Gloops} is not correct. Instead we
see that the square of diagram A in \fig{lo_gluon}, which at the
fixed-coupling order contributed
\begin{align}
  \frac{d \sigma_A}{d^2 k_T \, dy} \, = \, \frac{2 \, \as^3 \,
    C_F}{\pi^2} \, \int \frac{d^2 q}{({\bm q}^2)^2} \, \sum_{\lambda}
  \, \bigg| \frac{{\bm \epsilon}_\lambda \cdot ({\bm k} - {\bm
      q})}{({\bm k} - {\bm q})^2} \bigg|^2
\end{align}
to the gluon production cross section (see \eq{lip1} above), with the
help of Eqs.\ \peq{overall} and \peq{Gam} gets modified to
\begin{align}\label{A2}
  \frac{d \sigma_A}{d^2 k_T \, dy} \, = \, \frac{2 \, C_F}{\pi^2} \,
  \as \left( \Lambda_\text{coll}^2 \, e^{-5/3} \right) \, \int
  \frac{d^2 q}{({\bm q}^2)^2} \, \frac{\as^2 \left( {\bm q}^2 \,
      e^{-5/3} \right) \, \as^2 \left( ({\bm k} - {\bm q})^2 \,
      e^{-5/3} \right)}{\amu^2} \nn \\ \times \, \sum_{\lambda} \,
  \left| \frac{ {\bm \epsilon}_\lambda \cdot ({\bm k} - {\bm q}) \,
      \left( 1 + \amu \, \beta_2 \, L_a \right) - {\bm
        \epsilon}_\lambda \cdot {\bm k} \ \amu \, \beta_2 \,
      L_b}{({\bm k} - {\bm q})^2} \right|^2,
\end{align}
where $L_a$ and $L_b$ are defined in \eq{La} and \eq{Lb},
respectively.  It is clear that \eq{A2} can be written in terms of the
physical running-coupling constants, eliminating the $\mu$-dependence
completely.  However, we will postpone this last step until the next
Subsection, in which we will collect all the diagrams.

\FIGURE{\includegraphics[width=14cm]{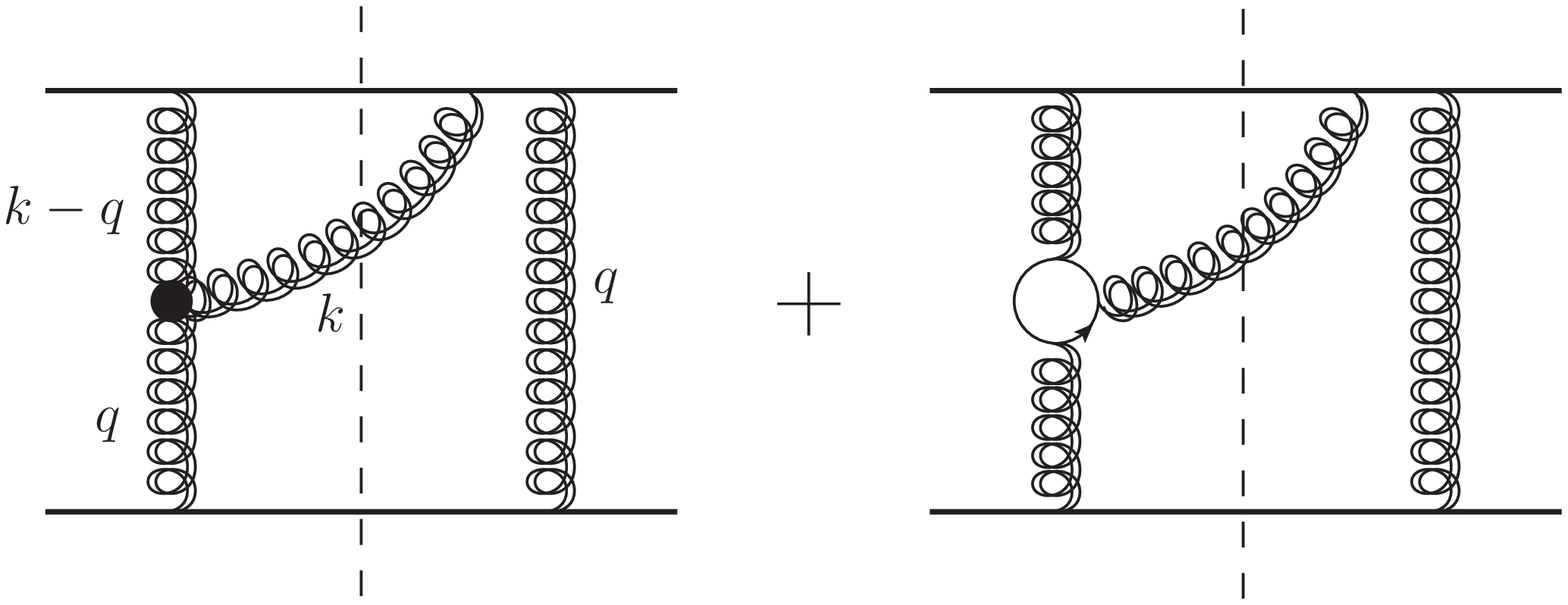}
\caption{\label{interf}
  Quark loop corrections to the interference terms between diagrams A
  and B from \fig{lo_gluon}. The notation is the same as in
  \fig{3Gloops}.}  }

Before we proceed, let us say a few words about the interference graphs
between the triple-gluon vertex diagram A and the bremsstrahlung
diagrams B and C in \fig{lo_gluon}. The quark loop corrections
resumming powers of $\amu \, N_f$ for such interference graphs are
illustrated in \fig{interf}. As was shown above in the discussion of
the diagrams in \fig{3Gloops}, cuts through the quark loop correction
to the triple-gluon vertex do not lead to collinearly-singular
contributions (i.e. contributions that survive in the
$\Lambda_\text{coll}^2 \rightarrow 0$ limit) and should be discarded.
Therefore the treatment of the quark loop corrections for the outgoing
gluon propagator should be the same for the interference graphs in
\fig{interf}, as it was for diagram A squared in \fig{3Gloops} and also for
the diagrams B and C squared in \fig{Bloops}: in each of the cases we
obtain a factor of $\as \left( \Lambda_\text{coll}^2 \, e^{-5/3}
\right)$. The rest of the quark loop corrections is included at the
amplitude level, as follows from the above analysis. For instance, the
right-hand side of each graph in \fig{interf} would bring in $\as
\left({\bm q}^2 \, e^{-5/3} \right)$, while the left-hand side would
give a contribution proportional to 
\begin{align}
  \frac{\as \left( {\bm q} \, e^{-5/3} \right) \, \as \left( ({\bm k}
      - {\bm q})^2 \, e^{-5/3} \right)}{\amu} \, \frac{ {\bm
      \epsilon}_\lambda \cdot ({\bm k} - {\bm q}) \, \left( 1 + \amu
      \, \beta_2 \, L_a \right) - {\bm \epsilon}_\lambda \cdot {\bm k}
    \ \amu \, \beta_2 \, L_b}{({\bm k} - {\bm q})^2}
\end{align}
as can be inferred from the above discussion and from \eq{A2}. 

We are now ready to combine all the diagrams together to determine the
scales of all the factors of the running coupling.


\subsection{The Result}
\label{ans}

Adding the contributions of the ``dressed'' bremsstrahlung diagrams
squared with the help of \eq{brem_rc} to the ``dressed'' diagram A
squared in \eq{A2}, and including all the interference diagrams we
finally write the expression for the gluon production cross section
with the running coupling corrections included:
\begin{align}\label{incl_rc3}
  & \frac{d \sigma}{d^2 k_T \, dy} \, = \, \frac{2 \, C_F}{\pi^2} \,
  \as \left( \Lambda_\text{coll}^2 \, e^{-5/3} \right) \, \int
  \frac{d^2 q}{({\bm q}^2)^2} \, \frac{\as^2 \left( {\bm q}^2 \,
      e^{-5/3} \right) \, \as^2 \left( ({\bm k} - {\bm q})^2 \,
      e^{-5/3} \right)}{\amu^2} \nn \\ & \times \, \sum_{\lambda} \,
  \left| \frac{ {\bm \epsilon}_\lambda \cdot ({\bm k} - {\bm q}) \,
      \left( 1 + \amu \, \beta_2 \, L_a \right) - {\bm
        \epsilon}_\lambda \cdot {\bm k} \ \amu \, \beta_2 \,
      L_b}{({\bm k} - {\bm q})^2} - \frac{{\bm \epsilon}_\lambda \cdot
      {\bm k}}{{\bm k}^2} \, \left( 1 + \amu \, \beta_2 \, \ln
      \frac{({\bm k} - {\bm q})^2 \, e^{-5/3}}{\msbar^2} \right)
  \right|^2,
\end{align}
where, again, $L_a$ and $L_b$ are defined by \eq{La} and \eq{Lb},
respectively.  Equation \peq{incl_rc3} can be written more compactly
as
\begin{align}\label{incl_rc4}
  \frac{d \sigma}{d^2 k_T \, dy} \, = \, \frac{2 \, C_F}{\pi^2} \, \as
  \left( \Lambda_\text{coll}^2 \, e^{-5/3} \right) \, \int \frac{d^2
    q}{({\bm q}^2)^2} \, \frac{\as^2 \left( {\bm q}^2 \, e^{-5/3}
    \right) \, \as^2 \left( ({\bm k} - {\bm q})^2 \, e^{-5/3}
    \right)}{\amu^2} \nn \\ \times \, \left[ \frac{{\bm k} - {\bm
        q}}{({\bm k} - {\bm q})^2} \, \left( 1 + \amu \, \beta_2 \,
      \ln \frac{Q_1^2 \, e^{-5/3}}{\msbar^2} \right) - \frac{{\bm
        k}}{{\bm k}^2} \, \left( 1 + \amu \, \beta_2 \, \ln
      \frac{Q_2^2 \, e^{-5/3}}{\msbar^2} \right) \right]^2
\end{align}
if we define momentum scales $Q_1$ and $Q_2$ by
\begin{align}\label{Q1}
  \ln \frac{Q_1^2}{\msbar^2} \, = \, \frac{({\bm k} - {\bm q})^2 \,
    \ln \frac{({\bm k} - {\bm q})^2}{\msbar^2} - {\bm q}^2 \, \ln
    \frac{{\bm q}^2}{\msbar^2}}{({\bm k} - {\bm q})^2 - {\bm q}^2} -
  \frac{{\bm q}^2 \, ({\bm k} - {\bm q})^2 \, {\bm k}^2}{ \left[ ({\bm
        k} - {\bm q})^2 - {\bm q}^2 \right]^3} \, \ln \frac{({\bm k} -
    {\bm q})^2}{{\bm q}^2}  + \frac{{\bm k}^2 \, \left[ ({\bm k}
      - {\bm q})^2 + {\bm q}^2 \right] }{2 \, \left[ ({\bm k} - {\bm
        q})^2 - {\bm q}^2 \right]^2}
\end{align}
and
\begin{align}\label{Q2}
  \ln \frac{Q_2^2}{\msbar^2} \, = \, \ln \frac{({\bm k} - {\bm
      q})^2}{\msbar^2} + \frac{{\bm k}^2}{({\bm k} - {\bm q})^2} \,
  \Bigg[ \frac{{\bm q}^2 \, ({\bm k} - {\bm q})^2 \, \left[ {\bm q}^2
      - ({\bm k} - {\bm q})^2 - 2 \, {\bm k}^2 \right]}{2 \, \left[
      ({\bm k} - {\bm q})^2 - {\bm q}^2 \right]^3} \, \ln \frac{({\bm
      k} - {\bm q})^2}{{\bm q}^2} \nn \\ + \frac{{\bm q}^2 \, \left[
      ({\bm k} - {\bm q})^2 - {\bm q}^2 \right] + {\bm k}^2 \, \left[
      ({\bm k} - {\bm q})^2 + {\bm q}^2 \right]}{2 \, \left[ ({\bm k}
      - {\bm q})^2 - {\bm q}^2 \right]^2} \Bigg].
\end{align}
We would like to point out that, as expected, the scales $Q_1$ and
$Q_2$ are independent of $\msbar$, as follows from Eqs.\ \peq{Q1} and
\peq{Q2}.

Recasting \eq{incl_rc4} in terms of physical running couplings we get
\begin{align}\label{incl_rc5}
  \frac{d \sigma}{d^2 k_T \, dy} \, = \, \frac{2 \, C_F}{\pi^2} \, \as
  \left( \Lambda_\text{coll}^2 \, e^{-5/3} \right) \, \int \frac{d^2
    q}{({\bm q}^2)^2} \, \as^2 \left( {\bm q}^2 \, e^{-5/3} \right) \,
  \as^2 \left( ({\bm k} - {\bm q})^2 \, e^{-5/3} \right) \nn \\ \times
  \, \left[ \frac{{\bm k} - {\bm q}}{({\bm k} - {\bm q})^2} \,
    \frac{1}{\as \left( Q_1^2 \, e^{-5/3} \right)} - \frac{{\bm
        k}}{{\bm k}^2} \, \frac{1}{\as \left( Q_2^2 \, e^{-5/3}
      \right)} \right]^2
\end{align}
which, together with Eqs.\ \peq{Q1} and \peq{Q2} can be considered the
final answer for the gluon production cross-section with the running
coupling corrections included. However, \eq{incl_rc5} is somewhat
unsatisfactory.  First, it does not explicitly exhibit the symmetry of
the problem under the ${\bm q} \leftrightarrow {\bm k} - {\bm q}$
interchange. Such symmetry follows from the $+ \leftrightarrow -$
interchange symmetry of the high-energy scattering of identical
hadrons/nuclei. We have broken this up-down symmetry in the diagrams
by choosing the $A^+ =0$ gauge: however, the physical answer should be
indeed independent of the gauge choice, and should be symmetric under
${\bm q} \leftrightarrow {\bm k} - {\bm q}$. While \eq{incl_rc5} does
posses such symmetry, it is not explicitly apparent in the form it is
written. Second of all, \eq{incl_rc5} does not look like the fixed
coupling cross-section \peq{lip2} multiplied by the factors of running
coupling, which is what one would expect from the procedure of setting
the scales of the running coupling constants.

Both of these problems are remedied when, after considerable algebra,
\eq{incl_rc5} can be recast into the following form:
\begin{align}\label{rc_incl}
  \frac{d \sigma}{d^2 k_T \, dy} \, = \, \frac{2 \, C_F}{\pi^2} \,
  \frac{ \as \left( \Lambda_\text{coll}^2 \, e^{-5/3} \right)}{{\bm
      k}^2} \, \int \frac{d^2 q}{{\bm q}^2 \, ({\bm k} - {\bm q})^2}
  \, \frac{\as^2 \left( {\bm q}^2 \, e^{-5/3} \right) \, \as^2 \left(
      ({\bm k} - {\bm q})^2 \, e^{-5/3} \right)}{\as \left( Q^2 \,
      e^{-5/3} \right) \, \as \left( Q^{* \, 2}\, e^{-5/3} \right)}
\end{align}
with the $\msbar$-independent momentum scale $Q$ defined by
\begin{align}\label{Qscale}
  \ln \frac{Q^2}{\msbar^2} \, & = \, \frac{1}{2} \, \ln \frac{{\bm
      q}^2 \, ({\bm k} - {\bm q})^2}{\msbar^4} - \frac{1}{4 \, {\bm
      q}^2 \, ({\bm k} - {\bm q})^2 \, \left[ ({\bm k} - {\bm q})^2 -
      {\bm q}^2 \right]^6} \, \Bigg\{ {\bm k}^2 \, \left[ ({\bm k} -
    {\bm q})^2 - {\bm q}^2 \right]^3 \notag \\[5pt]
  & \quad \; \times \, \bigg\{ \left[ \left[({\bm k}-{\bm
        q})^2\right]^2 - \left({\bm q}^2\right)^2 \right] \, \left[
    \left({\bm k}^2\right)^2 + \left(({\bm k}-{\bm q})^2 - {\bm
        q}^2\right)^2 \right] + 2 \, {\bm k}^2 \, \left[ \left({\bm
        q}^2\right)^3 - \left[({\bm k}-{\bm q})^2\right]^3
  \right] \notag \\[5pt]
  & \quad \; - {\bm q}^2 \, ({\bm k} - {\bm q})^2 \left[ 2 \,
    \left({\bm k}^2\right)^2 + 3 \, \left[({\bm k} - {\bm q})^2 - {\bm
        q}^2\right]^2 - 3 \, {\bm k}^2 \, \left[({\bm k} - {\bm q})^2
      + {\bm q}^2\right] \right] \, \ln
  \left( \frac{({\bm k} - {\bm q})^2}{{\bm q}^2} \right) \bigg\} \notag \\[5pt]
  & + \, i \, \left[ ({\bm k} - {\bm q})^2 - {\bm q}^2 \right]^3 \, \,
  \bigg\{ {\bm k}^2 \, \left[ ({\bm k} - {\bm q})^2 - {\bm q}^2\right]
  \, \left[ {\bm k}^2 \, \left[ ({\bm k} - {\bm q})^2 + {\bm
        q}^2\right] - \left({\bm q}^2\right)^2 -
    \left[({\bm k}-{\bm q})^2\right]^2 \right] \notag \\[5pt]
  & \quad \; + {\bm q}^2 \, ({\bm k} - {\bm q})^2 \, \left( {\bm k}^2
    \, \left[ ({\bm k} - {\bm q})^2 + {\bm q}^2\right] - 2 \,
    \left({\bm k}^2\right)^2 - 2 \, \left[ ({\bm k} - {\bm q})^2 -
      {\bm q}^2\right]^2 \right) \, \ln \left( \frac{({\bm k} - {\bm
        q})^2}{{\bm q}^2} \right) \bigg\} \notag \\[5pt]
  & \quad \; \times \, \sqrt{2 \, {\bm q}^2 \, ({\bm k} - {\bm q})^2 +
    2 \, {\bm k}^2 \, ({\bm k} - {\bm q})^2 + 2 \, {\bm q}^2 \, {\bm
      k}^2 - \left({\bm k}^2\right)^2 - \left({\bm q}^2\right)^2 -
    \left[({\bm k}-{\bm q})^2\right]^2} \Bigg\}.
\end{align}
Note that the scale $Q^2$ is complex-valued! The expression under the
square root in \eq{Qscale} is non-negative: therefore, to obtain the
complex conjugate scale $Q^*$ from \eq{Qscale} one only needs to
change the sign in front of the factor of $i$ in it.  The
cross-section \peq{rc_incl} is, of course, real, as it contains a
complex-valued coupling constant multiplied by its conjugate, $\as
\left( Q^2 \, e^{-5/3} \right) \, \as \left( Q^{* \, 2}\, e^{-5/3}
\right)$.

The scale $Q^2$ is explicitly symmetric under the ${\bm q}
\leftrightarrow {\bm k} - {\bm q}$ interchange: therefore, the cross
section \peq{rc_incl} is also symmetric under ${\bm q} \leftrightarrow
{\bm k} - {\bm q}$, as expected from the symmetries of this high
energy scattering problem.  Also \eq{rc_incl} clearly looks like the
fixed-coupling cross-section \peq{lip2} with three factors of
fixed-coupling replaced by the seven running couplings: we will refer
to this structure as {\sl the septumvirate} of couplings. 

\eq{rc_incl}, along with \eq{Qscale}, are the main results of this
work. They provide us with the gluon production cross section in high
energy quark--quark scattering with the running coupling corrections
included. 

Let us point out one interesting feature of our result \peq{rc_incl}.
It is well-known that the ${\bm q}$-integral in \eq{rc_incl} is
dominated by regions where either $\bm q \approx 0$ or $\bm q \approx
{\bm k}$ \cite{Gribov:1984tu,Blaizot:1987nc}. Due to the ${\bm q}
\leftrightarrow {\bm k} - {\bm q}$ symmetry each of these regions
contributes equally, so we can concentrate on only one of them.
Choosing the $\bm q \approx 0$ region we see from \eq{Qscale} that in
the $\bm q \rightarrow 0$ limit
\begin{align}
  \ln \frac{Q^2}{\msbar^2} \, = \, \ln \frac{{\bm k}^2}{\msbar^2} +
  \frac{1}{2}.
\end{align}
(In fact it is easier to see this by starting from Eqs.\ \peq{Q1},
\peq{Q2}, and \peq{incl_rc5}.)  Neglecting for simplicity numerical
constants such as $e^{5/3}$ and $1/2$ we approximate \eq{rc_incl} by
\begin{align}\label{rc_incl_approx}
  \frac{d \sigma}{d^2 k_T \, dy} \, \approx \, \frac{4 \, C_F}{\pi} \,
  \frac{ \as \left( \Lambda_\text{coll}^2 \right)}{({\bm k}^2)^2} \,
  \int\limits^{{\bm k}^2} \, \frac{d {\bm q}^2}{{\bm q}^2} \, \as^2
  \left( {\bm q}^2 \right).
\end{align}
The integral over ${\bm q}^2$ in \eq{rc_incl_approx} is cut off in the
IR by the saturation scale in the case of hadron--hadron,
hadron--nucleus, and nucleus--nucleus scattering. \eq{rc_incl_approx}
demonstrates that, loosely speaking, the exact \eq{rc_incl} can be
interpreted as having only two powers of $\as$ on top of $\as \left(
  \Lambda_\text{coll}^2 \right)$, with the two factors of $\as$
running at the smaller of the two scales ${\bm q}^2$ and $({\bm k} -
{\bm q})^2$.

Imagine that the quark--quark scattering considered here happens
within a larger process of hadronic (or nuclear) scattering.  After
integrating over ${\bm q}^2$ with the saturation scale $Q_s^2$ as the
IR cutoff, \eq{rc_incl_approx} becomes
\begin{align}\label{rc_incl_approx2}
  \frac{d \sigma}{d^2 k_T \, dy} \, \approx \, \frac{4 \, C_F}{\pi} \,
  \frac{\as \left( \Lambda_\text{coll}^2 \right) \, \as ({\bm k}^2) \,
    \as (Q_s^2)}{\left( {\bm k}^2 \right)^2} \, \ln \frac{{\bm
      k}^2}{Q_s^2},
\end{align}
which has a symmetric structure of one coupling running at the `gluon
resolution scale' $\Lambda_\text{coll}^2$, another at the IR cutoff
$Q_s^2$ characterizing the distribution functions, and the third
coupling running at the high transverse momentum scale ${\bm k}^2$.


\section{Discussion}
\label{disc}


\subsection{Conjecture for the Running Coupling Corrected 
$k_T$-Factorization Formula}
\label{conj}

Above in \eq{rc_incl} we have calculated running coupling corrections
to the lowest-order gluon production cross section in high-energy
quark--quark scattering. Our result can be easily generalized to the
case of nucleus--nucleus scattering in the McLerran--Venugopalan (MV)
model, in which each nucleon in the nuclei is assumed to be made of
valence quarks only. The generalization is accomplished by multiplying
the right-hand side of \eq{rc_incl} by $(N_c \, A_1) \, (N_c \, A_2)$,
with $A_1$ and $A_2$ the atomic numbers of the two nuclei. Such a
generalization would only be valid for large momenta $k_\perp = |{\bm
  k}| \gg Q_{s \, 1}, Q_{s \, 2}$ where $Q_{s \, 1}, Q_{s \, 2}$ are
the saturation scales of the two nuclei. In this regime non-linear
saturation effects are not yet important and can be neglected.

It would be useful though to generalize our result \peq{rc_incl} (i)
beyond the MV model, i.e., to include small-$x$ evolution
\cite{Bal-Lip,Kuraev:1977fs,Mueller:1994rr} in it and (ii) inside the
saturation region, where the nonlinear effects are important
\cite{Balitsky:1996ub,
  Balitsky:1997mk,Balitsky:1998ya,Kovchegov:1999yj, Kovchegov:1999ua,
  Jalilian-Marian:1997jx, Jalilian-Marian:1997gr,
  Jalilian-Marian:1997dw, Jalilian-Marian:1998cb, Kovner:2000pt,
  Weigert:2000gi, Iancu:2000hn, Ferreiro:2001qy}. Even for the fixed
coupling constant result for the gluon production cross section
including the small-$x$ evolution of \cite{Balitsky:1996ub,
  Balitsky:1997mk,Balitsky:1998ya,Kovchegov:1999yj, Kovchegov:1999ua,
  Jalilian-Marian:1997jx, Jalilian-Marian:1997gr,
  Jalilian-Marian:1997dw, Jalilian-Marian:1998cb, Kovner:2000pt,
  Weigert:2000gi, Iancu:2000hn, Ferreiro:2001qy} only exists for the
case of proton--nucleus scattering, i.e., for the case when $Q_{s}^A =
Q_{s \, 2} \gg Q_{s \, 1} = Q_s^p$ \cite{Kovchegov:2001sc}; Eqs.\ 
\peq{ktfact}, \peq{ktglueA}, and \peq{ktgluep}, which give the gluon
production in proton-nucleus scattering, were shown to be valid in
\cite{Kovchegov:2001sc} only for $k_\perp \gg Q_s^p$. Still it would
be very useful to generalize Eqs.\ \peq{ktfact}, \peq{ktglueA}, and
\peq{ktgluep} to include running coupling corrections.

To construct our conjecture for this generalization, let us start
working in the framework of the MV model. The running-coupling
corrections to the Glauber--Mueller formula \cite{Mueller:1989st}, the
forward amplitude of a $q\bar q$ dipole on a nucleus in the MV model,
was constructed in the Appendix of \cite{Kovchegov:2007vf} and is
given by Eq.\ (A8) there.  Generalizing this result to the case of a
$GG$ dipole scattering on a nucleus we write
\begin{align}\label{N0rc}
  N_G ({\bm r}, {\bm b}, y=0) \, = \, 1 - \exp \left[ - \pi \, \as
    \left( \frac{1}{{\bm r}^2} \right) \, \as \left( \Lambda^2 \right)
    \, \rho \, T({\bm b}) \, {\bm r}^2 \, \ln \frac{1}{|\bm r| \,
      \Lambda} \right].
\end{align}
(Strictly speaking the BLM analysis used in \cite{Kovchegov:2007vf} to
obtain the quark dipole amplitude has to be modified for the gluon
dipole: however, \eq{N0rc} is valid at least in the large-$N_c$ limit
when it could be obtained from Eq.~(A8) of \cite{Kovchegov:2007vf}
with the help of \eq{2NN} above.) The dipole amplitude given by the
Glauber--Mueller formula is rapidity-independent \cite{Mueller:1989st}
and serves as the initial condition for the subsequent BK/JIMWLK
evolution: this is denoted by putting $y=0$ in \eq{N0rc}. Above,
$\rho$ is the number density of nucleons in the nucleus and $T({\bm
  b})$ is the nuclear profile function equal to the length of the
nuclear medium at the impact parameter $\bm b$, such that $T({\bm b})
= 2 \, \sqrt{R^2 - b^2}$ for a spherical nucleus of radius $R$.
$\Lambda$ is some scale characterizing the nucleon, which, in general,
is non-perturbative and is comparable to $\Lambda_{QCD}$.

\eq{N0rc} is an approximation of a more exact expression
\cite{Kovchegov:2007vf}
\begin{align}\label{N0}
  N_G ({\bm r}, {\bm b}, y=0) \, = \, 1 - \exp \left( - \frac{1}{2} \,
    \rho \, T({\bm b}) \, \sigma^{GGN} ({\bm r}) \right)
\end{align}
with $\sigma^{GGN} ({\bm r})$ the cross-section of the gluon dipole
scattering on a nucleon calculated at the two-gluon exchange level
with the running coupling corrections \cite{Kovchegov:2007vf}
\begin{align}\label{sig_rc}
  \sigma^{GGN} ({\bm r}) \, = \, \int \frac{d^2 l_\perp}{[{\bm
      l}^2]^2} \, \as^2 \left( {\bm l}^2 \, e^{-5/3} \right) \, \left(
    2 - e^{i {\bm l} \cdot {\bm x}} - e^{- i {\bm l} \cdot {\bm x}}
  \right).
\end{align}
$\Lambda$ is the IR cutoff for the $\bm l$-integral in \eq{sig_rc}.
To construct the unintegrated gluon distribution at the lowest order
we expand \eq{N0} to the lowest non-trivial order in $\sigma^{GGN}$
and use the result along with \eq{sig_rc} in \eq{ktglueA} to obtain
\begin{align}\label{LOunint}
  \amu \, \phi_A^{LO} ({\bm k}, y=0) \, = \, N_c \, A \, \frac{\as^2
    \left( {\bm k}^2 \, e^{-5/3} \right)}{\pi} \, \frac{1}{{\bm k}^2}.
\end{align}
In arriving at \eq{LOunint} we have, for simplicity, assumed that the
nucleus is a cylinder of radius $R$ with the axis parallel to the beam
direction and $T({\bm b}) = 2 \, R$, such that we could replace $\rho
\, T({\bm b}) \rightarrow N_c \, A/(\pi \, R^2)$ with $A$ the atomic
number of the nucleus.

\eq{LOunint} demonstrates one important point: it shows that the
product $\amu \, \phi ({\bm k}, y)$ can be rewritten in terms of
running couplings. It may be possible to define the unintegrated gluon
distribution $\phi ({\bm k}, y)$ in such a way that it would be
expressible in terms of running couplings all by itself. However,
\eq{rc_incl} appears to suggest that such a separation is not necessary,
as it contains two powers of $\as \left( {\bm q}^2 \right)$ which are
likely to be absorbed into one distribution function and two powers of
$\as \left( ({\bm k} - {\bm q})^2 \right)$ which are likely to enter
the other distribution function.

Defining 
\begin{align}
  {\overline \phi} ({\bm k}, y) \, = \, \as \, \phi ({\bm k}, y)
\end{align}
we use \eq{LOunint} to rewrite \eq{rc_incl} (multiplied by $A \,
N_c^2$ for the $pA$ case) in the form of a generalization of
\eq{ktfact}:
\begin{align}\label{rc_LOkt}
  \frac{d \sigma}{d^2 k_T \, dy} \, = \, \frac{2 \, C_F}{\pi^2} \,
  \frac{1}{{\bm k}^2} \, \int d^2 q \ {\overline \phi}_p^{LO} ({\bm
    q}, 0) \, {\overline \phi}_A^{LO} ({\bm k} - {\bm q}, 0) \,
  \frac{\as \left( \Lambda_\text{coll}^2 \, e^{-5/3} \right)}{\as
    \left( Q^2 \, e^{-5/3} \right) \, \as \left( Q^{* \, 2}\, e^{-5/3}
    \right)}.
\end{align}
Again it seems natural that all factorized ${\bm q}$-dependence enters
one distribution function ${\overline \phi}_p^{LO} ({\bm q}, 0)$,
while all factorized $({\bm k} - {\bm q})$-dependence enters the other
distribution function ${\overline \phi}_A^{LO} ({\bm k} - {\bm q},
0)$. The terms depending on momenta ${\bm q}$ and ${\bm k} - {\bm q}$
in a way that can not be factorized into separate ${\bm q}$- and
$({\bm k} - {\bm q})$-dependent parts are all included in the scale
$Q$, which generates a ``vertex correction''.

\eq{rc_LOkt} is still an exact result as long as all momenta involved
are much larger than the saturation scale of the nucleus. However,
inspired by this formula we would like to conjecture the following
running-coupling generalization of \eq{ktfact}:
\begin{align}\label{rc_fact}
  \frac{d \sigma}{d^2 k_T \, dy} \, = \, \frac{2 \, C_F}{\pi^2} \,
  \frac{1}{{\bm k}^2} \, \int d^2 q \ {\overline \phi}_p ({\bm q}, y)
  \, {\overline \phi}_A ({\bm k} - {\bm q}, Y-y) \, \frac{\as \left(
      \Lambda_\text{coll}^2 \, e^{-5/3} \right)}{\as \left( Q^2 \,
      e^{-5/3} \right) \, \as \left( Q^{* \, 2}\, e^{-5/3} \right)}
\end{align}
with the distribution functions defined by
\begin{align}\label{rc_ktglueA}
  {\overline \phi}_A ({\bm k}, y) \, = \, \frac{C_F}{(2 \pi)^3} \,
  \int d^2 b \, d^2 r \, e^{- i {\bm k} \cdot {\bm r}} \ \nabla^2_r \,
  N_G ({\bm r}, {\bm b}, y)
\end{align}
and 
\begin{align}\label{rc_ktgluep}
  {\overline \phi}_p ({\bm k}, y) \, = \, \frac{C_F}{(2 \pi)^3} \,
  \int d^2 b \, d^2 r \, e^{- i {\bm k} \cdot {\bm r}} \ \nabla^2_r \,
  n_G ({\bm r}, {\bm b}, y).
\end{align}
Here $N_G$ should be found from the running-coupling BK/JIMWLK
evolution \cite{Balitsky:2006wa,Kovchegov:2006vj}, while $n_G$ should
be obtained from the running coupling BFKL equation
\cite{Kovchegov:2006wf}. The scale $Q$ is given by \eq{Qscale}. We
hope that \eq{rc_fact} is valid in the same regime as the original
\eq{ktfact}: it should be true both inside and outside the nuclear
saturation region, with and without the non-linear small-$x$
evolution. Of course only exact calculations can prove or disprove the
ansatz of \eq{rc_fact}.

Just like with the BK/JIMWLK nonlinear evolution equations, it is
likely that even if \eq{rc_fact} is valid, obtaining it may require a
subtraction similar to the one done in
\cite{Balitsky:2006wa,Kovchegov:2006vj} for the running-coupling BK
and JIMWLK equations, possibly leading, in the end, to an additive
UV-finite correction.  However, similar to the case of BK/JIMWLK
equations, it may be that for a choice of subtraction which preserves
the linear evolution in the running-coupling part of the answer, the
additive correction would be small
\cite{Albacete:2007yr,Balitsky:2006wa}. Since \eq{rc_fact} is likely
to correctly take into account all the linear running-coupling BFKL
evolution, it may still be a good approximation for the exact answer,
even if there is an additive correction to it.


\subsection{Multiplicity per Unit Rapidity}
\label{dNde}

Another interesting question to explore is the effect of our main
result \peq{rc_incl} on the integrated gluon multiplicity per unit
rapidity $dN/dy$. The latter is related to the hadronic multiplicity
in $pp$, $pA$ and $AA$ collisions, as was utilized in
\cite{Kharzeev:2000ph,Kharzeev:2001gp,Kharzeev:2001yq,Albacete:2007sm}.
Indeed to find $dN/dy$ one has to integrate \eq{rc_incl} over all
${\bm k}$: at the same time \eq{rc_incl} is only valid for $k_\perp
\gg Q_s$. However, we will use the trick which is valid for the
fixed-coupling calculations at least in the quasi-classical limit of
the MV model \cite{Lappi:2007ku,Blaizot:2010kh,Kovchegov:2000hz}: we
will integrate \peq{rc_incl} over $k_\perp$ with the saturation scale
$Q_s$ providing the IR cutoff. The result would be proportional to the
correct value of $dN/dy$ up to a constant. We have to note that the
fact that this procedure gives the correct answer for $dN/dy$ (up to a
constant) does not, in general, guarantee the same in the running
coupling case. In this sense our calculation below should be treated
as an approximation. 

The integration of the exact \eq{rc_incl} over $\bm k$ appears to be
rather daunting: instead we will use the approximate formula
\peq{rc_incl_approx} and write
\begin{align}\label{dNdy1}
  \frac{dN}{dy} \, \approx \, 4 \, C_F \, \frac{A^2}{S_\perp} \, \as
  \left( \Lambda_\text{coll}^2 \right) \, \int\limits_{Q_s^2}^\infty
  \, \frac{d {\bm k}^2}{({\bm k}^2)^2} \, \int\limits_{Q_s^2}^{{\bm
      k}^2} \, \frac{d {\bm q}^2}{{\bm q}^2} \, \as^2 \left( {\bm q}^2
  \right).
\end{align}
We assume that both nuclei are identical, each with atomic number $A$:
to generalize \peq{rc_incl_approx} to the case of $AA$ scattering we
multiplied it by $A^2$ (for simplicity we left out the number of
valence quarks in each nucleon, which would have given us an extra
factor of $N_c^2$). Here $S_\perp = \pi \, R^2$ is the cross-sectional
area of the nucleus, which is again assumed to be cylindrical.
Performing the integrals in \eq{dNdy1} yields
\begin{align}\label{dNdy2}
  \frac{dN}{dy} \, \approx \, 4 \, C_F \, S_\perp \, \left(
    \frac{A}{S_\perp} \right)^2 \, \frac{\as \left(
      \Lambda_\text{coll}^2 \right) \, \as^2 \left( Q_s^2
    \right)}{Q_s^2}.
\end{align}

To further evaluate \eq{dNdy2}, with the goal of determining the
$A$-dependence of $dN/dy$, one has to define the saturation scale
$Q_s$. We are working in the quasi-classical framework of the MV
model.  Therefore, the $A$-dependence of $Q_s$ that we are going to
find is necessarily limited to the quasi-classical regime, and is
going to be strongly modified by small-$x$ evolution with the running
coupling corrections, as shown in
\cite{Mueller:2003bz,Albacete:2004gw,Albacete:2007yr}. Even within the
classical approximation there appear to be two different ways of
defining $Q_s$:
\begin{itemize}
\item[(i)] Defining $Q_s$ by requiring that the dipole amplitude
  \peq{N0rc} is of the order of one at $|{\bm r}| = 1/Q_s$, or, more
  precisely, that the exponent in \eq{N0rc} is equal to $-1/4$ at
  $|{\bm r}| = 1/Q_s$, yields
  \begin{align}\label{Qs1}
    Q_s^2 \, = \, 4 \, \pi \, \as \left( Q_s^2 \right) \, \as \left(
      \Lambda^2 \right) \, \frac{A}{S_\perp} \, \ln
    \frac{Q_s}{\Lambda}
  \end{align}
  where we again replaced $\rho \, T({\bm b}) \rightarrow A/S_\perp$.
  Using \eq{Qs1} we eliminate factors of $A/S_\perp$ in \eq{dNdy2}
  obtaining
  \begin{align}\label{dNdy3}
    \frac{dN}{dy} \, \approx \, \frac{C_F}{4 \, \pi^2} \, S_\perp \,
    \frac{Q_s^2}{\ln^2 \frac{Q_s}{\Lambda}} \, \frac{\as \left(
        \Lambda_\text{coll}^2 \right)}{\as^2 \left( \Lambda^2 \right)}
    \ \propto \ \frac{A}{\ln^2 A}.
  \end{align}
  The proportionality follows from the assumption that for large
  enough $Q_s$ we may neglect the differences between $\Lambda$ and
  $\Lambda_{QCD}$ in \eq{Qs1} and write $Q_s^2 \propto A^{1/3}$.
  Identifying the number of participating nucleons $N_{part}$ with $A$
  we see that the multiplicity per participant $(1/N_{part}) \,
  dN/dy$, in this naive model with cylindrical nuclei, is a decreasing
  function of $A$. It seems that running coupling effects alone are
  not sufficient to describe RHIC heavy ion collision data on hadron
  multiplicity per participant along the lines of
  \cite{Kharzeev:2001yq}. Other effects used in
  \cite{Kharzeev:2001yq}, like realistic nuclear profiles leading to
  two different saturation scales $Q_s^{max}$ and $Q_s^{min}$ for the
  two nuclei appear to be more important in describing the data (see
  \cite{Drescher:2006pi,Kuhlman:2006qp}).
  
\item[(ii)] Another, possibly less justified way of defining $Q_s$ is
  to require that 
  \begin{align}\label{Qsunint}
    {\overline \phi} (|{\bm k}| = Q_s, y) \, = \, \frac{S_\perp}{(2 \,
      \pi)^2}.
  \end{align}
  This is motivated by the saturation requirement for the unintegrated
  gluon distribution in the IR, though we have to point out that the
  distributions entering the $k_T$-factorization formula \peq{ktfact}
  given by Eqs.\ \peq{ktglueA} and \peq{ktgluep} in fact go to zero in
  the $k_T \rightarrow 0$ limit \cite{Braun:2000bh,Kharzeev:2003wz}.
  \eq{Qsunint} gives
  \begin{align}\label{Qs2}
    Q_s^2 \, = \, 4 \, \pi \, \as^2 \left( Q_s^2 \right) \,
    \frac{A}{S_\perp},
  \end{align}
  which, when used in \eq{dNdy2} yields
\begin{align}\label{dNdy4}
  \frac{dN}{dy} \, \approx \, \frac{C_F}{4 \, \pi^2} \, S_\perp \,
  \frac{Q_s^2}{\as^2 \left( Q_s^2 \right)} \, \as \left(
    \Lambda_\text{coll}^2 \right) \ \propto \ A.
  \end{align}
  We see that in this scenario the particle multiplicity per
  participant does not change with centrality, reinforcing our earlier
  conclusion that running coupling corrections alone are not enough to
  describe the centrality dependence of the multiplicity per participant
  observed in heavy ion collisions at RHIC.
\end{itemize}

Indeed our above conclusions are limited to the quasi-classical regime
and are derived under the assumption that \eq{dNdy2}, which gives us
the multiplicity of particles with $k_\perp > Q_s$, correctly
describes the centrality-dependence of the total particle
multiplicity. More detailed (possibly numerical) work is needed to
derive the centrality dependence of the conjecture \peq{rc_fact}.
Another way of including an $A$ dependence in Eqs.\ \peq{dNdy3} and
\peq{dNdy4} is to identify $\Lambda_\text{coll}$ with $Q_s$ (or make
it proportional to some other power of $Q_s$). However, it is not
clear whether this is justified theoretically, since $Q_s$ is clearly
not the smallest momentum scale in the problem in the full non-linear
regime.  From the phenomenological side such a substitution would
introduce an (extra) decrease of $(1/N_{part}) \, dN/dy$ with $A$ in
both Eqs.\ \peq{dNdy3} and \peq{dNdy4}, making a successful comparison
with RHIC data more difficult to achieve.


\subsection{Summary}
\label{summary}

To summarize, in this paper we have found the running coupling
corrections to the lowest order gluon production cross section in
hadronic and nuclear high energy scattering. Our exact results are
presented in Eqs. \peq{rc_incl} and \peq{Qscale}. An approximate
simplified version of the exact formula is given in
\eq{rc_incl_approx2}. Based on the results \peq{rc_incl} and
\peq{Qscale} we have conjectured the $k_T$-factorization formula for
the gluon production cross section with running coupling corrections
included, given in Eqs. \peq{rc_fact}, \peq{rc_ktglueA}, and
\peq{rc_ktgluep}. We hope future work will verify our conjecture.


\acknowledgments

Yu.K. is grateful to Genya Levin and Larry McLerran for discussions.
Yu.K. would also like to thank the organizers of the workshop on High
Energy Strong Interactions 2010 at the Yukawa Institute for
Theoretical Physics, Kyoto, Japan, where part of this work was
completed.

This research is sponsored in part by the U.S. Department of Energy
under Grant No. DE-SC0004286.


\appendix

\renewcommand{\theequation}{A\arabic{equation}}
  \setcounter{equation}{0}
\section{Loop integral evaluation}
\label{A}

The goal of this Appendix is to evaluate the expression in \eq{loop2}.
Following the standard procedure for evaluating loop integrals in
dimensional regularization \cite{Peskin:1995ev,Sterman:1994ce} we
introduce Feynman parameters $x$ and $y$ and shift the integration
variable by defining
\begin{align}
  {\tilde l}^\mu \, = \, l^\mu - y \, q^\mu + x \, (k-q)^\mu.
\end{align}
After dropping the tilde we rewrite \eq{loop2} as
\begin{align}\label{Gam1}
  \Gamma \, & = \, i \, g^3 \, f^{abc} \, N_f \, \epsilon^\lambda_\mu
  (k) \, (k-q)_\nu^\perp \, \int\limits_0^1 dx \, \int\limits_0^{1-x}
  dy \, \int \frac{d^d l}{(2 \, \pi)^d} \, \frac{\Tr \left[ \gamma^\mu
      \, \gamma^\alpha \, \gamma^\nu_\perp \, \gamma^\beta \, \gamma^+
      \, \gamma^\delta \right]}{\left[ l^2 - (1-x-y) \, \left( y \,
        {\bm
          q}^2 + x \, ({\bm k} - {\bm q})^2 \right) \right]^3} \nn \\
  & \times \left\{ l_\alpha \, l_\beta \, \left[ - {\bar y} \,
      q_\delta - x \, (k-q)_\delta \right] + l_\alpha \, l_\delta \,
    \left[ y \, q_\beta - x \, (k-q)_\beta \right] + l_\beta \,
    l_\delta \, \left[ y \, q_\alpha + {\bar x} \, (k-q)_\alpha
    \right] \nn \right. \\ & \left. + \left[ y \, q_\alpha + {\bar x}
      \, (k-q)_\alpha \right] \, \left[ y \, q_\beta - x \,
      (k-q)_\beta \right] \, \left[ - {\bar y} \, q_\delta - x \,
      (k-q)_\delta \right] \right\}
\end{align}
with 
\begin{align}
  {\bar x} \, = \, 1-x, \ \ \ {\bar y} \, = \, 1-y
\end{align}
and $\alpha$, $\beta$, and $\delta$ some internal indices. 

The first three terms in the curly brackets of \eq{Gam1} are evaluated
by substituting
\begin{align}
  l_\alpha \, l_\beta \, \rightarrow \, \frac{1}{d} \ l^2 \,
  g_{\alpha\beta}
\end{align}
which makes the Dirac traces trivial. Denoting the contribution of
these first three terms by $\Gamma_3$ we perform the Wick rotation and
integrate over $l$ to obtain
\begin{align}\label{Gam31}
  \Gamma_3 \, = - g \, f^{abc} \, k^+ \, {\bm \epsilon}_\lambda \cdot
  ({\bm k} - {\bm q}) \, \frac{\amu \, N_f}{2 \, \pi} \,
  \int\limits_0^1 dx \, (1+x) \int\limits_0^{1-x} dy \, \left\{ \ln
    \left[ \frac{(1-x-y) \, \left( y \, {\bm q}^2 + x \, ({\bm k} -
          {\bm q})^2 \right)}{\msbar^2} \right] + 1 \right\}
\end{align}
where we have added the $N_f$ piece of the triple-gluon vertex
counterterm in the $\overline {\text MS}$ renormalization scheme.
Integrating \peq{Gam31} over $x$ and $y$ and completing $N_f$ to the
full beta-function using \peq{repl} yields
\begin{align}\label{Gam32}
  \Gamma_3 \, = \, 2 \, g \, f^{abc} \, k^+ \, \amu \, \beta_2 \ {\bm
    \epsilon}_\lambda \cdot ({\bm k} - {\bm q}) \, \Bigg\{ \frac{1}{12
    \, \left[ ({\bm k} - {\bm q})^2 - {\bm q}^2 \right]^2} \, \Bigg[
  \left( 19 \, {\bm q}^2 - 16 \, ({\bm k} - {\bm q})^2 \right) \,
  \left( ({\bm k} - {\bm q})^2 - {\bm q}^2 \right) \nn \\ + 3 \, ({\bm
    k} - {\bm q})^2 \, \left( 4 \, ({\bm k} - {\bm q})^2 - 5 \, {\bm
      q}^2 \right) \, \ln \frac{({\bm k} - {\bm q})^2}{\msbar^2} + 3
  \, {\bm q}^2 \, \left( 4 \, {\bm q}^2 - 3 \, ({\bm k} - {\bm q})^2
  \right) \, \ln \frac{{\bm q}^2}{\msbar^2} \Bigg] - \frac{7}{12}
  \Bigg\}.
\end{align}

We now turn our attention to the last term in the curly brackets of
\eq{Gam1}. We denote the contribution of this term by $\Gamma_4$.
Wick rotation, integration over $l$, and evaluation of the
trace of Dirac matrices, after somewhat convoluted algebra, gives
\begin{align}\label{Gam41}
  \Gamma_4 \, & = \, g \, f^{abc} \, k^+ \, \epsilon^\lambda_\mu (k)
  \, (k-q)_\nu^\perp \ \frac{\amu \, N_f}{2 \, \pi} \, \int\limits_0^1
  dx \, \int\limits_0^{1-x} dy \ \frac{1}{y \, {\bm q}^2 + x \, ({\bm
      k} - {\bm q})^2} \nn \\ & \times \, \bigg\{ g^{\mu\nu} \, \left[
    x^2 \, ({\bm k} - {\bm q})^2 + y \, (1+x) \, {\bm q}^2 \right] - 2
  \, x \, (1 - 2\, x - 2 \, y) \ q^\mu \, q^\nu_\perp + 2 \, x \, (1-2
  \, x) \ q^\mu \, k^\nu_\perp \bigg\}.
\end{align}
Integrating over $x$ and $y$ in \peq{Gam41}, replacing $N_f$ with the
help of \peq{repl}, and bringing $\epsilon^\lambda_\mu (k) \,
(k-q)_\nu^\perp$ inside the curly brackets we obtain
\begin{align}\label{Gam42}
  \Gamma_4 \, = \, 2 \, g \, f^{abc} \, k^+ \, \amu \, \beta_2 \ 
  \Bigg\{ {\bm \epsilon}_\lambda \cdot ({\bm k} - {\bm q}) \, \left[
    \frac{3 \, ({\bm k} - {\bm q})^2 \, \left( {\bm q}^2 - ({\bm k} -
        {\bm q})^2 + {\bm q}^2 \, \ln \frac{({\bm k} - {\bm
            q})^2}{{\bm q}^2} \right)}{4 \, \left[ ({\bm k} - {\bm
          q})^2 - {\bm q}^2 \right]^2} + 1 \right] \nn \\ + \, {\bm
    \epsilon}_\lambda \cdot {\bm q} \ ({\bm k} - {\bm q}) \cdot {\bm
    q} \ \frac{{\bm q}^2 - ({\bm k} - {\bm q})^2 + {\bm q}^2 \, \ln
    \frac{({\bm k} - {\bm q})^2}{{\bm q}^2}}{2 \, \left[ ({\bm k} -
      {\bm q})^2 - {\bm q}^2 \right]^2} \nn \\ - {\bm
    \epsilon}_\lambda \cdot {\bm q} \ ({\bm k} - {\bm q}) \cdot {\bm
    k} \ \frac{\left[ ({\bm k} - {\bm q})^2 - {\bm q}^2 \right] \,
    \left[ ({\bm k} - {\bm q})^2 + 3 \, {\bm q}^2 \right] - {\bm q}^2
    \, \left[ 3 \, ({\bm k} - {\bm q})^2 + {\bm q}^2 \right] \, \ln
    \frac{({\bm k} - {\bm q})^2}{{\bm q}^2}}{2 \, \left[ ({\bm k} -
      {\bm q})^2 - {\bm q}^2 \right]^3} \Bigg\}.
\end{align}

In order to more efficiently combine Eqs.\ \peq{Gam32} and \peq{Gam42}
we note that the triple gluon vertex in the leading-order diagram A in
\fig{lo_gluon} contributes
\begin{align}\label{3Glue}
  - 2 \, g \, f^{abc} \, k^+ \, g^{\mu\nu}
\end{align}
such that, when multiplied by $\epsilon^\lambda_\mu (k) \,
(k-q)_\nu^\perp$, it becomes
\begin{align}\label{3G1}
  2 \, g \, f^{abc} \, k^+ \, {\bm \epsilon}_\lambda \cdot ({\bm k} -
  {\bm q}).
\end{align}
Therefore, defining 
\begin{align}\label{GLO}
  \Gamma_\text{LO} \, = \, 2 \, g \, f^{abc} \, k^+
\end{align}
we finally write for $\Gamma = \Gamma_3 + \Gamma_4$
\begin{align}\label{Gam}
  \Gamma \, = \, \Gamma_\text{LO} \, \amu \, \beta_2 \ \Bigg\{ {\bm
    \epsilon}_\lambda \cdot ({\bm k} - {\bm q}) \, L_a - {\bm
    \epsilon}_\lambda \cdot {\bm k} \, L_b \Bigg\},
\end{align}
where we define
\begin{align}\label{La}
  L_a \, \equiv \,  \frac{({\bm k} - {\bm q})^2 \, \ln \frac{({\bm k}
      - {\bm q})^2 \, e^{-5/3}}{\msbar^2} - {\bm q}^2 \, \ln
    \frac{{\bm q}^2 \, e^{-5/3}}{\msbar^2}}{({\bm k} - {\bm q})^2 -
    {\bm q}^2} - \frac{{\bm q}^2 \, ({\bm k} - {\bm q})^2 \, {\bm
      k}^2}{ \left[ ({\bm k} - {\bm q})^2 - {\bm q}^2 \right]^3} \,
  \ln \frac{({\bm k} - {\bm q})^2}{{\bm q}^2}  + \frac{{\bm
      k}^2 \, \left[ ({\bm k} - {\bm q})^2 + {\bm q}^2 \right] }{2 \,
    \left[ ({\bm k} - {\bm q})^2 - {\bm q}^2 \right]^2}
\end{align}
and
\begin{align}\label{Lb}
  L_b \equiv \frac{{\bm q}^2 \, ({\bm k} - {\bm q})^2 \, \left[ {\bm
        q}^2 - ({\bm k} - {\bm q})^2 - 2 \, {\bm k}^2 \right]}{2 \,
    \left[ ({\bm k} - {\bm q})^2 - {\bm q}^2 \right]^3} \, \ln
  \frac{({\bm k} - {\bm q})^2}{{\bm q}^2} + \frac{{\bm q}^2 \, \left[
      ({\bm k} - {\bm q})^2 - {\bm q}^2 \right] + {\bm k}^2 \, \left[
      ({\bm k} - {\bm q})^2 + {\bm q}^2 \right]}{2 \, \left[ ({\bm k}
      - {\bm q})^2 - {\bm q}^2 \right]^2}.
\end{align}
Note that the first term on the right-hand side of \eq{La} is formally
similar to what was obtained for the kernel of the running-coupling
BK/JIMWLK evolution equations (see Eq.\ (86) in
\cite{Kovchegov:2006vj}). However the similarity does not extend
beyond this term.



\providecommand{\href}[2]{#2}\begingroup\raggedright\endgroup


\end{document}